\documentclass[prx,superscriptaddress,aps, onecolumn, reprint, showpacs, longbibliography, footnoteinbib]{revtex4-1}

\usepackage{graphicx}
\usepackage{amssymb}   
\usepackage{amsfonts}
\usepackage{amsmath}
\usepackage{dsfont}
\usepackage{natbib}
\usepackage{dcolumn}   
\usepackage{bm}        
\usepackage[mathscr]{eucal}
\usepackage[dvipsnames]{xcolor}
\usepackage[colorlinks, linkcolor=OrangeRed,citecolor=RoyalBlue,urlcolor=NavyBlue]{hyperref}
\usepackage[all]{hypcap} 
\usepackage{mathtools}
\usepackage{dsfont}

\definecolor{myblue}{rgb}{0,0,0.75}

\newcommand{\ep}{\varepsilon}
\newcommand{\IPR}{\text{IPR}}

\newcommand{\be}{\begin{equation}}
\newcommand{\ee}{\end{equation}}
\def\ba{\begin{aligned}}
\def\ea{\end{aligned}}
\newcommand{\bea}{\begin{eqnarray}}
\newcommand{\eea}{\end{eqnarray}}
\def\bes{\begin{subequations}}
\def\ees{\end{subequations}}
\def\bal{\begin{align}}
\def\eal{\end{align}}

\newcommand{\la}{\left\langle}
\newcommand{\ra}{\right\rangle}
\newcommand{\lv}{\left|}
\newcommand{\rv}{\right|}
\newcommand{\lb}{\left[}
\newcommand{\rb}{\right]}
\newcommand{\lp}{\left(}
\newcommand{\rp}{\right)}

\newcommand\lrp[1]{\lp#1\rp}
\newcommand\lrb[1]{\lb#1\rb}
\newcommand\lrv[1]{\lv#1\rv}
\newcommand\lra[1]{\la#1\ra}

\newcommand\bra[1]{\ensuremath{\la#1\rv}}
\newcommand\ket[1]{\ensuremath{\lv#1\ra}}
\newcommand\braket[2]{\ensuremath{\lra{#1 | #2}}}
\newcommand\bracket[3]{\ensuremath{\lra{#1 \lrv{#2} #3}}}
\newcommand\mean[1]{\ensuremath{\overline{#1}}}

\renewcommand{\hat}[1]{{\widehat #1}}

\usepackage{soul}
\usepackage{ulem} 

\usepackage{graphicx}
\usepackage{amsmath,amssymb,bm}
\usepackage{enumerate}
\usepackage{braket}        

\usepackage{amsmath}

\begin{document}


\title{Rare thermal bubbles at the many-body localization transition from the Fock space point of view}
 \author{Giuseppe De Tomasi}
\affiliation{T.C.M. Group, Cavendish Laboratory, JJ Thomson Avenue, Cambridge CB3 0HE, United Kingdom}
\author{Ivan M. Khaymovich}
\affiliation{Max-Planck-Institut f\"ur Physik komplexer Systeme, N\"othnitzer Stra{\ss}e 38, 01187-Dresden, Germany}%
\author{Frank Pollmann}
\affiliation{Department of Physics, Technische Universit\"at M\"unchen, 85747 Garching, Germany}
\author{Simone Warzel}
\affiliation{Department of Mathematics, Technische Universit\"at M\"unchen, 85747 Garching, Germany}

\begin{abstract}{
In this work we study the many-body localization (MBL) transition and relate it to the eigenstate structure in the Fock space.
Besides the standard entanglement and multifractal probes, we introduce the radial probability distribution of eigenstate coefficients with respect to
the Hamming distance in the Fock space 
and relate the cumulants of this distribution to the properties of the quasi-local integrals of motion in the MBL phase.
We demonstrate non-self-averaging property of the many-body fractal dimension $D_q$ and directly relate it to the jump of $D_q$ as well as of the localization length of the integrals of motion at the MBL transition.
We provide an example of the continuous many-body transition confirming the above relation via the self-averaging of $D_q$ in the whole range of parameters.
Introducing a simple toy-model, which hosts ergodic thermal bubbles, we give analytical evidences both in
standard probes and in terms of newly introduced radial probability distribution
that the MBL transition in the Fock space is consistent with the avalanche mechanism for delocalization, i.e., the Kosterlitz-Thouless scenario.
Thus, we show that the MBL transition can been seen as a transition between ergodic states to non-ergodic extended states and put the upper bound for the disorder scaling for the genuine Anderson localization transition with respect to the non-interacting case.
}
\end{abstract}
\maketitle

\section{Introduction}
Understanding the emergence of ergodicity in closed quantum many-body systems is an active front of research~\cite{Deutsch1991,Srednicki1994,Srednicki1996,rigol2008thermalization,Polkon_2011,DAlessio2016ETH}.
Generic interacting systems are expected to thermalize under their own quantum dynamics. Nevertheless, thermalization may fail if the system is subjected to strongly quenched disorder, giving arise to a new phase of matter dubbed as many-body localized (MBL)~\cite{Basko06,gornyi2005interacting,Pal2010,oganesyan2007localization,ALET2018498, huse2015review, CollAba}.

The MBL phase is best understood in terms of an emergent form of integrability,
which is characterized by the existence of an extensive set of quasi-local conserved quantities, which strongly hinder thermalization in the system~\cite{serbyn2013local, huse2014phenomenology,ros2015integrals}.
As a consequence, the system has Poisson level statistic, area-law entanglement, and the partial local structure of the initial state is maintained under the evolution~\cite{Pal2010, oganesyan2007localization, Luitz15}.
Instead at weak disorder, the system is in an ergodic phase, meaning that Eigenstate Thermalization Hypothesis~\cite{DAlessio2016ETH, rigol2008thermalization, Srednicki1994, Deutsch1991} (ETH) holds, and therefore local observable thermalize. This implies that the system is fully described in terms of few macroscopic conserved quantities, i.e., energy and/or particles number.

\begin{figure}[h!]
    \includegraphics[width=1.\linewidth]{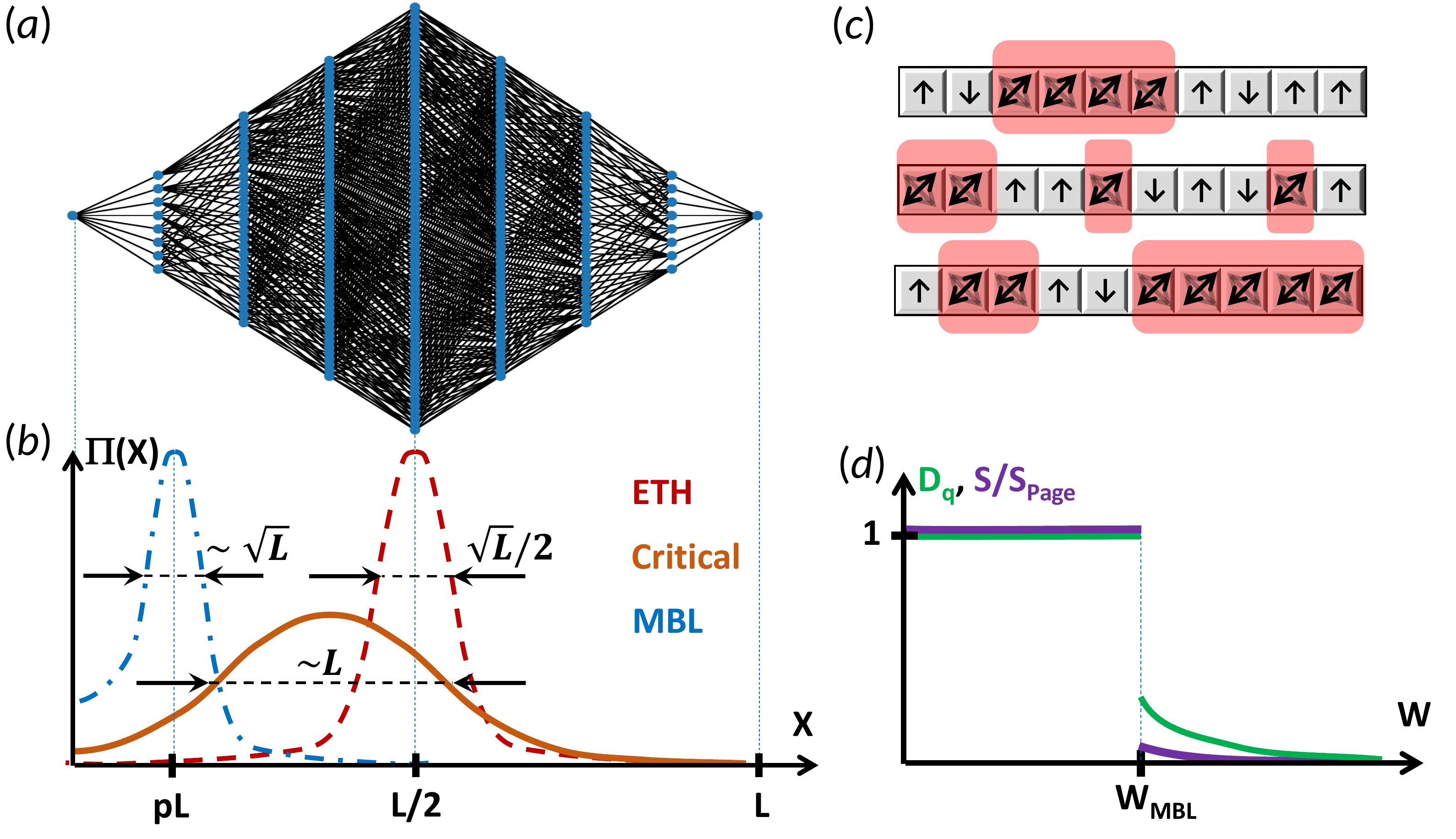}
  \caption{(a)~Representation in the Fock space (hypercube) of the many-body Hamiltonian $\hat H_{\text{MBL}}$ versus the Hamming distance. The blue nodes represent the basis vectors which are connected by $\hat H_{\text{MBL}}$. (b)~Cartoon picture of the radial probability distribution $\Pi(x$) of an eigenstate with respect to the same Hamming distance $x$ as in the panel (a) in the Fock space from the maximum of the wave function. In the ergodic/ETH phase $\Pi(x) = \frac{1}{2^L} \binom{L}{x}$ with the maximum at $x=L/2$, where most of the sites in the Fock space are. In the MBL phase, $\Pi(x)$ is skewed on the left with the width $\sim \sqrt{L}$. At the critical point, $\Pi(x)$ is much broader and fluctuations are extensive in $L$. (c)~Pictorial representation of the rare thermal bubbles (in red), lengths of which are highly fluctuating at the critical point. The spins in red regions are entangled to each other (shown by many arrows), while the spins away from these regions are considered to be frozen (up or down). (d)~Sketch of the jump of the entanglement entropy $S$ normalized by its ergodic value $S_{\text{Page}}$ and the fractal dimension $D_q$ across the MBL transition.}
    \label{fig:Car1}
\end{figure}

A quantum phase transition, referred to as MBL transition~\cite{Pal2010}, is believed to separate an ergodic phase from an MBL one.
The MBL transition is a dynamical phase transition, meaning that it occurs at the level of individual eigenstates even at high energy density.
In the last decade an enormous effort, both numerically and theoretically~\cite{Basko06, Canovi11, Kill14, Giu17, Luitz15, Var17, Mirlin2016dot, Vedi17, Vosk15, Bera15}, has been made to understand the nature of this transition. Nevertheless, only little is known about the MBL transition.
Numerically, the critical exponents associated with a putative second-order type of the transition, are in disagreement with generic bounds~\cite{Chandran_2015_bounds, Luitz15}.
This incongruence could be due to the fact that the system sizes analyzed
using exact diagonalization techniques are too small to capture the true asymptotic behavior.

Very recently, several theoretical works have doubted the underline assumptions that the transition is of the second order.
Phenomenological renormalization group studies suggest that the MBL transition could be of a Kosterlitz-Thouless (KT) type~\cite{DeRoeck2017_MBL_RG, DeRoeck2018_MBL_Avalanche, Goremykina2019-1, Goremykina2019-2, Alan_2019_1, Alan_2020_2,Alan_2019_3}.
This possible scenario is properly predicted by a possible delocalization mechanism called avalanche theory,
which takes into account non-perturbative effects possibly destabilizing the MBL phase~\cite{DeRoeck2017_MBL_RG, DeRoeck2018_MBL_Avalanche}.
Strictly speaking, the theory states that the presence of thermal bubbles in the system due to unavoidable entropic arguments is enough to destabilize the MBL phase
if the localization length $\xi_{\text{loc}}$ settled by the disorder strength exceeds a finite critical length. An immediate consequence of this mechanism
is that the MBL is characterized not by the {\it divergence} of the correlation length, as one expects from the ordinary second-order transition, but
by a {\it finite jump} of the inverse localization length across the transition~\cite{DeRoeck2018_MBL_Avalanche, Goremykina2019-2}.

A complementary interesting perspective is to characterize MBL systems in the Fock space, see Fig.~\ref{fig:Car1}~(a) for a pictorial representation.
This paradigm is based on the original idea of mapping a disordered quantum dot to a localization problem in the Fock space~\cite{Alt97} which has been developed further recently~\footnote{The main directions here can be characterized as the mapping of the MBL to the localization on hierarchical structures like
a random regular graph~(see, e.g., \cite{Biroli:2012vk,Deluca14,Alt16,TikMir16,Tik16,Biroli2017Dynamics, Sonner17,Lemarie17Small_K, Biroli2018delocalization,Kra18,parisi2019anderson,bera19,DeTomasi2019Subdiffusion,Tikh2019_K(w),Tikh2019Critical,Biroli_Tarzia2020subdiffusion,tikhonov2020eigenstate}),
considerations of the spin models with uncorrelated on-site disorder like a Quantum Random energy model (see, e.g.,~\cite{
Laumann2014QREM,Baldwin2016qRem,faoro2019QREM,Smelyanskiy2020QREM,Kechedzhi2018QREM, biroli2020QREM} and some other ones~~\cite{Logan19, Basko06, Roy1, Roy2, GDT_2020_Fock}),
as well as the associating of the MBL transition with the ergodic transition in random-matrix models~\cite{Kravtsov_NJP2015,Tarzia_2020,LNRP2020_RRG,LNRP2020_WE}.
}.
This has been used to provide evidence of the existence of an MBL transition by Basko, Aleiner, and Altshuler in their seminal work~\cite{Basko06}.
Ergodicity is then defined through the fractal dimensions $D_q$, which quantify the spread of a state in the Fock space~\cite{Evers2008Review}.
Ergodic states at infinite temperatures are believed to behave like random vectors~\cite{Page1993,WE_ETH-footnote}, therefore they are spread homogeneously over the entire Fock space and $D_q=1$.
Instead, non-ergodic states cover only a vanishing fraction of the Fock space, $0\le D_q<1$.
A genuine localization in the Fock space requires $D_q=0$, though due to the many-body nature of the problem is never reached at finite disorder~\cite{Luitz15,Tikhonov2018MBL_long-range,Mace_Laflorencie2019_XXZ,Luitz_Khaymovich_BarLev_multifrac_SciPost2020,Tarzia_2020}.
Thus, the MBL transition can be seen as an ergodic to non-ergodic transition in the Fock space, with $D_q=1$ in the ETH phase and $D_q<1$  in the MBL phase.

In a recent work~\cite{Mace_Laflorencie2019_XXZ}, the behavior of $D_q$ for a certain MBL model has been inspected using extensive numerical calculations. The MBL transition was found to be characterized by a jump in the fractal dimensions $D_q$ at the critical point.
The aforementioned investigations lead to the indication of the existence of an MBL transition. However, a clear connection between the above two viewpoints is still missing.

In this work, we focus on the existence of the above jump in the fractal dimensions and on its connection to the avalanche theory, i.e., to the KT-type transition from another perspective.
Based on the breakdown of self-averaging for $D_q$ at the transition and on the recently developed relation of $D_q$ to the entanglement entropy~\cite{DeTomasi_2020}, we show that the MBL transition is consistent with the jump of $D_q$ from $D_q = 1$ to $D_q<1/2$.
In addition, we focus on the radial distribution of a many-body eigenstate in the Fock space around its maximum,
relate it to the behavior of local integrals of motion~\cite{Imbrie2016}, and, thus, confirm the consistency of the KT-scenario~\cite{DeRoeck2018_MBL_Avalanche, Goremykina2019-2} for the MBL transition, see Fig.~\ref{fig:Car1} for an overall picture.

This paper is organized as follows. In Sec.~\ref{sec:Model} we introduce the model and the indicators that we inspected numerically.
In particular, we study the inverse participation ratio of eigenstate coefficients in the Fock space, from which we extract the fractal dimensions and the
radial probability of eigenstate coefficients.
Section~\ref{Sec:Results} represents the numerical results concerning the fractal dimensions and the radial probability distribution.
In Sec.~\ref{Sec:Bubble} we show our analytical considerations which underline the connection between the avalanche theory and the observed jump in the fractal dimensions.

In Sec.~\ref{Sec:continuous}, we provide an example of an non-interacting model with many-body filling,
which is known to have a delocalization-localization transition and characterized by a diverging localization length at this transition.
We show the main difference of this model from the MBL transition which is believed to have a discontinuity in the inverse localization length.
Finally, we draw our conclusions and outlooks in Sec.~\ref{Sec:Conclusion}.

\section{Model and Methods}\label{sec:Model}
We study the random quantum Ising model~\cite{Imbrie2016} with the Hamiltonian
\begin{equation}
\label{eq:Hamiltonian}
 \hat H_{\text{MBL}} = \sum_i^L \hat \sigma^x_i + \sum_i^L h_i \hat \sigma_i^z + V\sum_{i}^L J_i \hat \sigma_{i}^z \hat \sigma_{i+1}^z,
\end{equation}
of a spin chain of the length $L$ with periodic boundary conditions and the Pauli operators at site $i$ given by $\sigma^\alpha_i$ for $\alpha \in \{x,y,z\}$.
$h_i$ and $J_{i}$ are independent random variables uniformly distributed in $[-W,W]$ and $[0.8, 1.2]$,
respectively. $W$ is the disorder and $V = 1$ is the interaction strengths.

In~\cite{Imbrie2016} under mild assumptions of the absence on energy level attraction
the existence of the MBL phase has been established for sufficiently large, but finite $W$.
Moreover, numerically the critical disorder strength of the MBL transition has been identified as
$W_c\approx 3.5$~\cite{2019Abanin_many}. 

In the non-interacting limit of $V=0$ the Hamiltonian, Eq.~\eqref{eq:Hamiltonian}, represents a system of uncoupled spins, which is trivially localized in the sense that all eigenstates are product states.
In this limiting case, the Hamiltonian can be expressed, $\hat H = \sum_i \epsilon_i \hat \tau_i^z$, through its integrals of motion $\hat \tau_i^z = \hat U_i \hat \sigma_i^z \hat U_i^\dagger$ obtained from the original spins by a single-spin rotation
\begin{equation}\label{eq:U_mat}
\hat U_i = \begin{pmatrix} \cos \frac{\theta_i}{2}  & -\sin\frac{\theta_i}{2} \\ \sin\frac{\theta_i}{2} & \cos\frac{\theta_i}{2} \end{pmatrix} \ ,
\end{equation}
with $\sin \theta_i = 1/\sqrt{1+h_i^2}$ and the single-spin energies $\epsilon_i =  \sqrt{1+h_i^2}$.
This example directly shows that the eigenstates
$\{ \ket{\underline{\tau}^z} \}$ of $\hat H_{\text{MBL}}$ with $V=0$ are adiabatically connected to the $\sigma_z$-basis product states $\ket{\underline{\sigma}^z} = \otimes_i \ket{\sigma_i^z }$ with $\sigma_i^z\in \{-1,1\}$ through local rotations
$\prod_i \hat U_i  \ket{\underline{\sigma}^z} = \ket{\underline{\tau}^z}$.
In Ref.~\onlinecite{Imbrie2016} it has been shown that for $V\ne 0$ and sufficiently large $W$, $\hat H_{\text{MBL}}$ can still be
diagonalized via a sequence of local rotations which adiabatically connect the eigenstates to the product states in the $\sigma_z$-basis.

Another useful perspective of the model Eq.~\eqref{eq:Hamiltonian} is to consider it as an Anderson model on the Fock space.
For this, one can rewrite the Hamiltonian in the $\sigma_z$-basis and associate the first term in Eq.~\eqref{eq:Hamiltonian} to the hopping and the rest to the on-site correlated disorder on a $L$-dimensional hypercube,
\begin{equation}
\label{eq:Hamiltonian_fock}
 \hat H_{\text{MBL}} = \sum_{\underline{\sigma}^z \sim \underline{\sigma'}^z} \ket{\underline{\sigma}^z} \bra{\underline{\sigma'}^z} + \sum_{\underline{\sigma}^z } E_{\underline{\sigma}^z}
 \ket{\underline{\sigma}^z} \bra{\underline{\sigma}^z} \ ,
\end{equation}
where $\ket{\underline{\sigma}^z}$ stands for the configuration given by the vector $\underline{\sigma}^z$ of $L$ values $\sigma_i^z = \{+1,-1\}$, while
$\underline{\sigma}^z \sim \underline{\sigma'}^z$ means that the corresponding vectors differ by a single spin flip.
In this representation the first sum in $\hat H_{\text{MBL}}$ can be understood as the Laplace operator on the hypercube as it connects spins configurations, which differ by one spin flip,
and $E_{\underline{\sigma}^z} = \sum_i h_i \sigma_i^z + \sum_i J_{i} \sigma_i^z \sigma_{i+1}^z$ are the diagonal energies.
It is important to note that the $2^L$ diagonal entries $\{E_{\underline{\sigma}^z}\}$ are strongly correlated random variables since they are constructed only from $2L$ random variables $\{h_i\}$ and $\{J_i\}$.
Indeed, even though typical fluctuations of the entries scales as $\sqrt{L}$, their level spacings $E_{\underline{\sigma}^z} - E_{\underline{\sigma'}^z} $ are $ O(1)$ if
$\underline{\sigma}^z \sim \underline{\sigma'}^z$.

This model should be distinguished from the Quantum Random Energy Model (QREM)
different from its classical counterpart~\cite{Derrida1981REM, Gold91, Manai20}
for which the diagonal entries  $\{E_{\underline{\sigma}^z}\}$ are replaced by independent identically distributed Gaussian random variables $ \mathcal{N}(0, W^2 L)$
by a transverse field term $\sum_i^L \hat \sigma^x_i$.
The QREM has an Anderson transition at $W_c \sim \sqrt{L} \log{L}$ (cf.~Eq.~(10.15)~\cite{AizWar15}).

Ergodic properties of an eigenstate $\ket{E}$ of $\hat H_{\text{MBL}}$ in Eq.~\eqref{eq:Hamiltonian_fock} can be quantified using the generalized inverse participation ratio ($\IPR_q$)
\begin{equation}
 \IPR_q = \sum_{\underline{\sigma}^z}^{2^L} \lrv{\braket{\underline{\sigma}^z | E} }^{2q},
\end{equation}
which quantifies the spread of $\ket{E}$ over the Fock space. Through the $\IPR_q$ the fractal dimensions are defined as
\begin{equation}
\label{eq:D_q}
 D_q =  \frac{\log{\IPR_q}}{(1-q)L\log{2}}.
\end{equation}
Ergodic states at infinite temperature are characterized by $D_q=1$ since they extend over the entire Fock space $\lrv{\braket{\underline{\sigma}^z | E}}^{2} \sim 1/2^L$.
In general, $0<D_q<1$ corresponds to the non-ergodic (or multifractal) states, while the extreme case $D_q=0$ refers to the localized ones.

For a model similar to $\hat H_{\text{MBL}}$ considered in \cite{Mace_Laflorencie2019_XXZ} it has been shown that in the ergodic phase ($W<W_c$)
mid-spectrum eigenstates show $D_q =1$. Instead, in the MBL phase $D_q < 1$ and the fractal dimension experiences a jump at the critical point.
It is important to point out that due to the many-body nature of the wave-functions $D_q>0$~\cite{Luitz15,Tikhonov2018MBL_long-range,Mace_Laflorencie2019_XXZ,Tarzia_2020} for any finite values of $W$, even deeply in the MBL phase.

The last observation is closely related to the tensor product structure of the Fock space and the Hamiltonian's local structure.
Indeed, in the non-interacting limit ($V=0$) spins are decoupled, i.e., $\IPR_q^{V=0} = \prod_i^L \IPR_q^{(i)}$ and the one-site $\IPR_q^{(i)}$ is smaller than one
\begin{equation}\label{eq:IPR_single}
 \IPR_q^{(i)} =  \sum_{\sigma_z\in \{\uparrow, \downarrow \}} \lrv{\braket{\sigma_z | \tau_i^z} }^{2q} <1 \ .
\end{equation}
As a consequence, $\text{IPR}_q^{V=0} \sim 2^{-(q-1)\mean{D_q^0} L + \mathcal{O}(\sqrt{L}) }$ decays exponentially with the strictly positive exponent
\be\label{eq:D_q>0}
\mean{D_q^0} = \frac{\mean{\log \IPR_q^{(i)}}}{(1-q)\log 2} >0 \ .
\ee
This fractal exponent is self-averaging as it is the sum of independent random variables and has fluctuations $\mathcal{O}(1/\sqrt{L}) $ shown above in the exponent of the IPR~\cite{GDT_2020_Lea}.  
At this point, it is important to appreciate the difference between the MBL phase of
$\hat H_{\text{MBL}}$ and the localized phase for the QREM. The first one is characterized by a strictly positive fractal dimension, while in the second one $D_q=0$.

For a better understanding of the ergodicity properties from the Fock-space point of view, we define the radial probability distribution $\Pi(x)$~\footnote{The following properties hold $\Pi(x)\ge 0$ and $\sum_x \Pi(x)=1$.} of an eigenstate $\ket{E}$ as
\begin{equation}
 \Pi(x) = \mean{\sum_{d(\underline{\sigma}^z,\underline{\sigma}_0^z)=x} \lrv{ \braket{\underline{\sigma}^z | E}   }^2},
 \label{eq:Pi}
 \end{equation}
where the sum runs over the $\binom{L}{x}$ spin states $\{\ket{\underline{\sigma}^z}\}$ which differ by $x$ flips (i.e., at the Hamming distance $d(\underline{\sigma}^z,\underline{\sigma}_0^z)=x$) from $\ket{\underline{\sigma}_0^z}$,  which corresponds to the maximal eigenstate coefficient $\max_{\underline{\sigma}^z}  \lrv{\braket{\underline{\sigma}^z | E} }^2 =  \lrv{  \braket{\underline{\sigma}_0^z | E}   }^2$.
The overbar indicates the average over disorder and a few mid-spectrum eigenstates.

Compared to the $\IPR_q$, $\Pi(x)$ gives more information and is a good probe of the eigenstate's local
structure in the Fock space.  In particular, we can study the spread of $\Pi(x)$ by defining the moments
\begin{equation}
 \mean{X^n} = \sum_x x^n \Pi(x),
 \label{eq:X}
\end{equation}
and the mean-square displacement of it
\begin{equation}
 \Delta X^2 = \mean{X^2} - \mean{X}^2.
\label{eq:X_2}
 \end{equation}

In the ergodic phase, the infinite-temperature wave function is spread homogeneously on the Fock space and $\Pi^{\text{Erg.}}(x)$ 
is given by a binomial distribution,
\be\label{eq:P(r)_p}
\Pi_p\lrp{L,x} = \lrp{L\atop x}  (1-p) ^{L-x} p ^{x} \ ,
\ee
with $p=1/2$ and therefore $\mean{X} = L/2$ and $\Delta X^2 = L/4$.
In the opposite limit of a strongly localized system, which can be approximated by 
the non-interacting case ($V=0$), $\Pi(x)$ is still given by the binomial probability distribution Eq.~\eqref{eq:P(r)_p}, however the value of $p = \mean{\sin^2\lrp{{\theta_i}/2}}$ now strongly depends on the disorder $W$ as
\begin{equation}
\label{eq:p}
p =  \frac12 -\mean{ \frac{\lrv{h_i}}{2\sqrt{1+h_i^2}} } = \frac12 - \frac{\sqrt{W^2+1}-1}{2W}. \
\end{equation}
As expected, $p\simeq 1/2-{W}/{4}\to 1/2$ as $W\to 0$ like in the ergodic phase, but for $V=0$ it happens due to the system localization in $\sigma^x$-basis,
and $p \simeq 1/(2W)\to 0$ in the opposite limit of $W\to \infty$.

Note that the binomial probability distribution in Eq.~\eqref{eq:P(r)_p}, can be approximated
by a Gaussian distribution with mean and the variance given by
\be\label{eq:X,dX_2_Gauss}
\mean{X} = p L \ , \quad \Delta X^2 = p(1-p)L \ .
\ee

%
%

\section{Results} \label{Sec:Results}
In order to relate the ergodicity properties of the considered system with the local structure of its eigenstates in the Fock space, we focus on
the behavior of the radial probability distribution $\Pi(x)$ of mid-spectrum eigenstates of $\hat H_{\text{MBL}}$.
However, for sake of completeness we start our analysis by investigating some standard MBL indicators which quantify ergodicity in the real and Fock space, such as bipartite entanglement entropy and $\IPR_q$ and compare their properties.


\begin{figure}[t]
    \includegraphics[width=1.\linewidth]{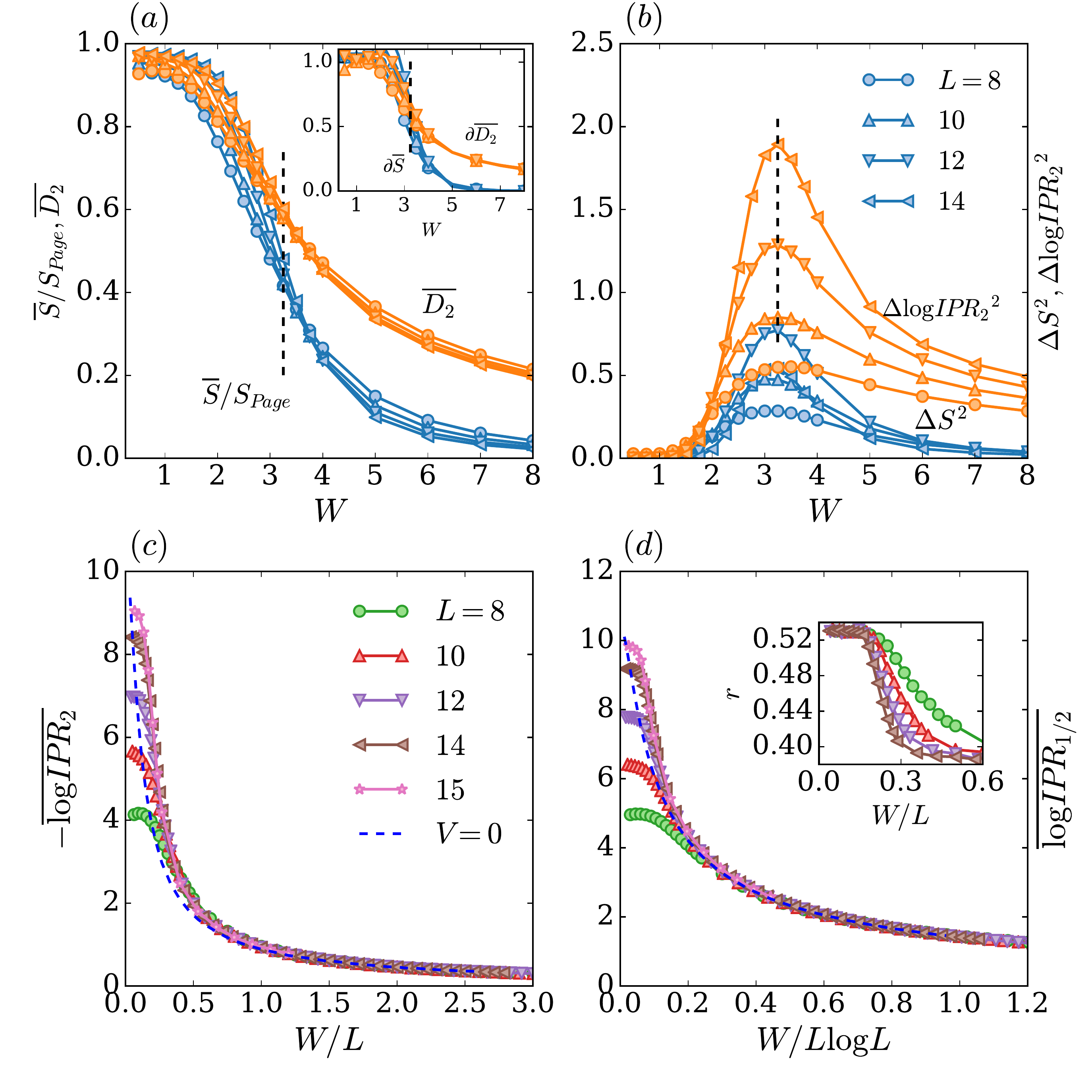}
    \caption{(a)~The averaged bipartite entanglement entropy rescaled by the Page value $S/S_{\text{Page}}$ (blue) and the averaged fractal dimension $D_2$ (orange) versus the disorder strength $W$ for $\hat H_{\text{MBL}}$. 
    Its inset shows the discrete derivatives of the entanglement entropy $\mean{\partial S} = 2\lrp{S(L)-S(L')}/(L-L')\log 2$~\cite{dS_footnote} (blue) and $\mean{\partial D_2} = -\log\lrp{\IPR_2(L)/\IPR_2(L')}/(L-L')\log 2$ (orange) with respect to the system size $L$.
    (b)~Variances of the entanglement entropy $S$ (blue) and of the $-\log{\IPR}_2$ (orange) versus $W$. In both panels the vertical dashed black line is a guide for eyes indicating the MBL transition.
    Symbols shown in panel (b) correspond to different system sizes.
    Next panels show disorder dependence of $-\mean{\log{\IPR}_q}$ for (c)~$q=2$ and (d)~$q=1/2$ with the correspondingly rescaled $W$ according to the non-interacting limit~\eqref{eq:D_q_non_int}.
    In both panels, the dashed blue line show the corresponding non-interacting case $V=0$.
    The inset in (d) shows the ratio $r$-statistics of the level spacings versus the rescaled $W$.}
    \label{fig:fig1}
\end{figure}
The entanglement entropy has been found to be a surrounding resource to test and quantify ergodicity in a system.

Figure~\ref{fig:fig1}(a) shows the half-chain bipartite entanglement entropy $S = -\text{Tr}[\rho_{L/2} \log \rho_{L/2}]$ (blue lines)
of the reduced density matrix $\rho_{L/2}$ of a mid-spectrum eigenstates of $\hat H_{\text{MBL}}$. 
As expected, at weak disorder $S$ shows the volume law, $S\sim L$, and flows towards the Page value $S_{\text{Page}} = L/2\log{2} - 1/2$~\cite{Page1993},
which is the value for states randomly drawn in Fock space.

Instead, at strong disorder $S$ has an area law scaling, $S\sim O(1)$, and thus $S/S_{\text{Page}}\sim 1/L$ tends to zero.
The crossover between the two behaviors occuring at $W_c\approx 3.5$ indicates the MBL transition.

The averaged fractal dimension $\overline{D}_2$ in Eq.~\eqref{eq:D_q} is also shown in Fig.~\ref{fig:fig1}~(a) (orange lines) and its behavior is consistent with the one of $S$.
In the ergodic phase $\mean{D_2}\approx 1$, while for $W>W_c$ the fractal dimension converges with $L$ to a value which is
strictly smaller than one ($\mean{D_2}<1$).

A few comments are in order:
the finite-size flow of $\mean{S}/S_{\text{Page}}$ and $\mean{D_2}$ to unity with the increasing system size within the ergodic phase
and the stability in the localized phase.
The discrete derivatives $\mean{\partial S} = 2\lrp{S(L)-S(L')}/(L-L')\log 2$~\cite{dS_footnote} and $\mean{\partial D_2} = -\log\lrp{\IPR_2(L)/\IPR_2(L')}/(L-L')\log 2$ in the
inset of Fig.~\ref{fig:fig1}(a) tend to converge to discontinuous functions of $W$ with increasing $L$, with
zero value of $\mean{\partial S}$ and strictly positive value of $\mean{\partial D_2}$ in the MBL phase.
In addition, as clearly seen from Fig.~\ref{fig:fig1}(a), both $\mean{S}/S_{\text{Page}}$ and $\mean{D_2}$ deviate from their ergodic values at the same disorder amplitude.
At the critical point the variance of the entanglement entropy  $\Delta S^2$, which has been shown to be a useful probe for the transition~\cite{Luitz15, Kill14},  diverges with $L$ (blue lines in Fig.~\ref{fig:fig1}(b)).
In analogy with $S$, we also inspect the variance of $-\log{IPR_2} = L D_2 \log{2}$, which also diverges around the critical point (orange lines in Fig.~\ref{fig:fig1}(b)).

The above observations give an indication of the following scenario.
The upper bound for the unaveraged $S\leq D_1 L \log 2$ derived in~\cite{DeTomasi_2020} and the deviation of $\mean{S}$
from its ergodic Page-value limit $\mean{S} =  S_{\text{Page}}$ at the same $W$ as $\mean{D_2}$, is consistent
with the jump in $\mean{D_q}$ across the MBL transition from unity in the ergodic phase to a certain positive value $D_q<1/2$ in the MBL phase.
Moreover, the divergent fluctuations of $D_q$ might lead to the saturation of the above bound~\cite{DeTomasi_2020} at the transition pushing $\mean{S}$ to undergo the volume-to-area law scaling transition and $\mean{D_q}$ to experience a jump.

\begin{figure}[t]
    \includegraphics[width=1.\linewidth]{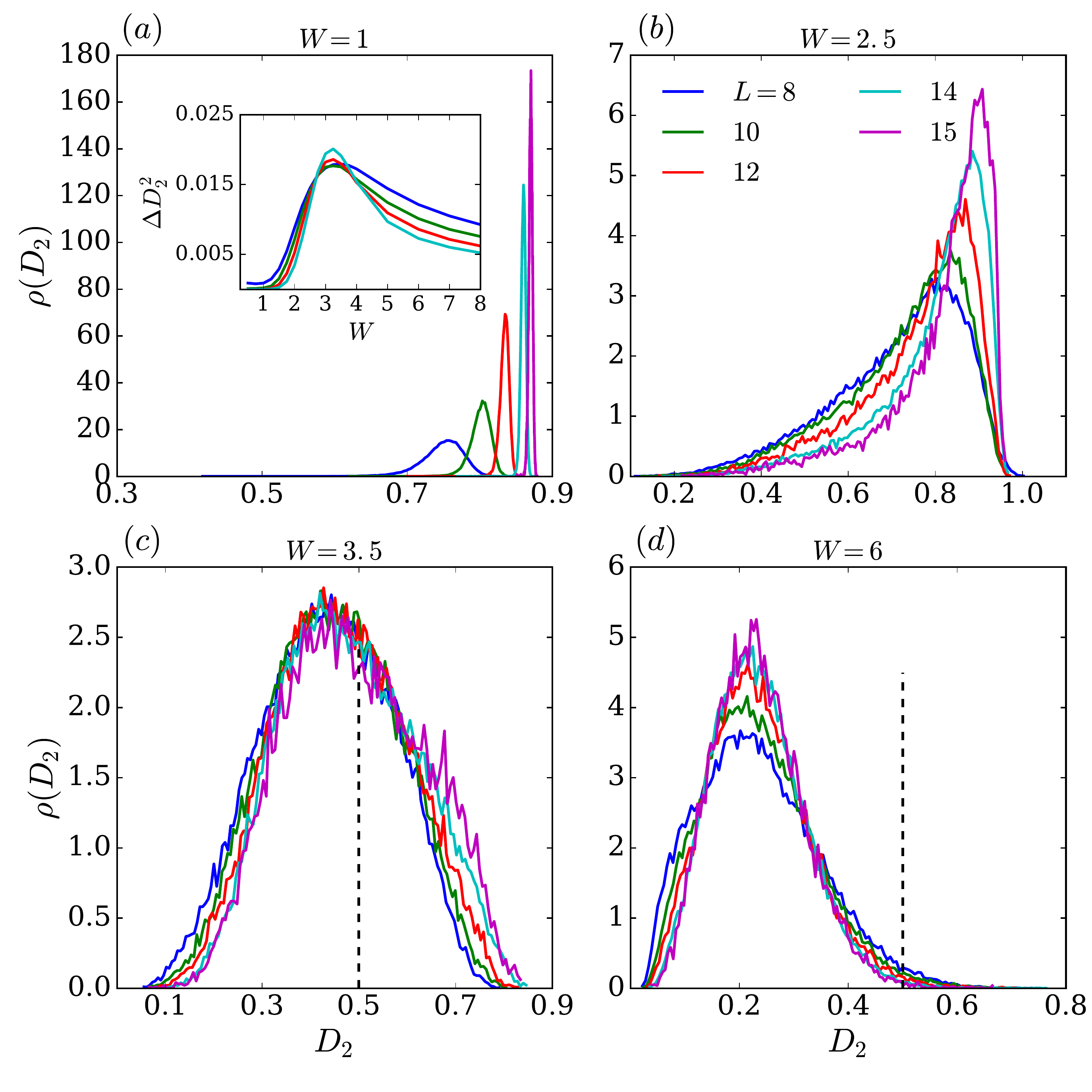}
    \caption{Probability distribution function $\rho$ of the fractal dimension $D_2$ for different disorder strength (a)~$W=1$, (b)~$W=2.5$, (c)~$W=3.25$, (d)~$W=6$.
    Different colors represent system sizes $L$ shown in the legend of panel (b).
    The inset shows the variance of the fractal dimension versus $W$ establishing the non-self averaging property close to the MBL transition ($\Delta D_2^2\sim O(L^0)$). The vertical black dashed line in (c) and (d) indicate the value $D_2=1/2$.}
    \label{fig:fig2}
\end{figure}

To obtain more insights in Fig.~\ref{fig:fig2} we analyze the behavior of the fractal dimension fluctuations via
the probability distribution of $D_2$ for several relevant values of $W$.
Deeply in the ergodic phase $D_2$ tends to unity and its fluctuations are exponentially suppressed with $L$, as dictated by ETH at infinite temperature (Fig.~\ref{fig:fig2}(a-b)). Instead, in the localized phase (Fig.~\ref{fig:fig2}(d)), we expect $\mean{D_2}<1/2$ and self-averaging in the sense that $\Delta D_2^2 \sim 1/L$ as shown in the inset of Fig.~\ref{fig:fig2}(a).
The self-averaging of $D_q$ in the MBL phase can be understood in its non-interacting limit ($V=0$),
\be\label{eq:dDq_2_self-av}
\Delta D_q^2 = \frac{\sum_i \Delta (\log{IPR_q^{(i)}})^2}{L^2 (1-q)^2 \log^2{2}} \sim \frac{1}{L} \ .
\ee
On the contrary, in the critical region ($W\approx W_c$), we find that $D_q$ is not self-averaging since its fluctuation do not decay with $L$ (inset of Fig.~\ref{fig:fig2})
and the probability distribution is stuck and not shrinking with increasing $L$ (Fig.~\ref{fig:fig2}(c)).

At critical points of single-particle problems, self-averaging is usually absent if $D_q$ demonstrates a jump~\cite{Evers2008Review}.
Its absence is far from being trivial in many-body problems and it provides another case for the jump of $D_q$ at the critical point.
Summarizing, at the MBL transition $D_q$ might be characterized by a jump and it is not self-averaging.
This non-self-averaging and the relation between $D_q$ and $S$~\cite{DeTomasi_2020} might drive the simultaneous jump-like transitions
in $\mean{S}$ and $\mean{D_q}$.

It is important to point out once again that $D_q$ is strictly positive due to the many-body nature of the problem.
As a consequence, the MBL transition cannot been considered as an Anderson-localization transition in the Fock space with $D_q = 0$ in the localized phase,
but rather as a transition between an ergodic ($D_q=1$) and a non-ergodic extended ($0<D_q<1$) phase.
In order to reach a genuine localization transition in the Fock space one needs to rescale the disorder strength with $L$.
Naively, as a first approximation, one might replace the Fock space on-site energies with independent distributed random variables with typical fluctuation $\sim \sqrt{L}$ as is done for the QREM~\cite{Laumann2014QREM, Baldwin2016qRem,faoro2019QREM,Smelyanskiy2020QREM,Kechedzhi2018QREM}.  If this would be the case, the Anderson transition would occur at $W_{AT} \sim \sqrt{L} \log{L}$ also for the quantum Ising model Eq.~\eqref{eq:Hamiltonian}.
Nevertheless, the Fock space on-site energies are strongly correlated and they cannot be approximated as independent random variables.
Due to the presence of these correlations (similar to~\cite{Aubry80}) we expect that stronger disorder is needed to localize many-body states in the Fock space.

In order, to understand the correct scaling of $W_{AT}$ with $L$, we rely on the exactly solvable non-interacting case (c$V=0$) providing the lower bound for $D_q$
and then check numerically if the same scaling works  as well at strong disorder in the interacting case.
This comparison is motivated by the belief~\cite{Imbrie2016} that interacting eigenstates in the MBL phase are adiabatically connected to the ones of a non-interacting problem.
Consequently, we expect to rescale $W$ with $L$ in {\it the same way} as in the non-interacting problem in order to have a genuine localization in the Fock space.
Straightforward calculations of the single-spin $\IPR_q^{(i)} = \sin^4\tfrac{\theta_i}2 +\cos^4\tfrac{\theta_i}2$ in the non-interacting model show that at large disorder $W$
\be
\mean{D_q^0} \sim
\begin{cases}
  W^{-2q}\quad &q<1/2,\\
  \frac{\log W}{W} \quad &q=1/2,\\
  W^{-1} \quad &q>1/2.
\end{cases}
\label{eq:D_q_non_int}
\ee
Thus rescaling $W \sim L$, we have $\IPR_q\sim O(1)$ for $q>1/2$ and the system exhibits localization
in the Fock space. In Fig.~\ref{fig:fig1}(c) we plot $-\mean{\log{\IPR_2}}$ for $V=1$ which is collapsed after the rescaling $W/L$ in agreement with the prediction of the non-interacting calculation.

The last observation should be compared to the scaling of $W$ for the QREM, where $W_{\text{QREM}}\sim \sqrt{L}\log{L} \ll L$ and the random many-body energies are uncorrelated.
However, due to the limitation of system sizes achievable with exact diagonalization techniques, it is important to notice that a reasonable collapse of the curves in Fig~\ref{fig:fig1}(c) can also be obtained  by rescaling $W/(\sqrt{L}\log{L})$.
Thus, in order to draw a distinction between the two scalings, we consider another moment $q=1/2$ for $\IPR_q$.
The critical value for the QREM is independent of $q$, while for the MBL model we expect a different scaling with $L$ for $q \leq 1/2$, as shown in Eq.~\eqref{eq:D_q_non_int}.
Figure~\ref{fig:fig1}(d) clearly demonstrates the collapse of $\mean{\log{\IPR_{1/2}}}$ for several $L$, by rescaling $W$ with $L\log{L}$ in complete agreement with the prediction of Eq.~\eqref{eq:D_q_non_int}.

Recently, some works~\cite{Lev19,Sels_2020} claim that the critical point for the MBL transition is $L$-dependent and shifts as $W_{c}\sim L$.
These claims are in conflict with our results for which the system is already Anderson-localized in the Fock space ($D_q=0$) provided $W\sim L$.
Localization in the Fock space is a much stronger breaking of ergodicity than the one defined using local observables (ETH).
Since we expect to have $D_q=1$ in a putative ergodic phase at infinite temperature,
it implies that the critical point of the MBL transition, if it scales with system size, would have to scale slower than $L$ ($W_c/L \rightarrow 0$), which rules out the prediction in Ref~\cite{Lev19, Sels_2020}.

Furthermore, a more refined analysis puts even stronger constraint on the possible behavior of $W_c$ with $L$ (if any). It is natural to expect that the stronger the interaction is, the stronger is the disorder needed to break the ergodicity and reach MBL.
From  this perspective, the uncorrelated on-site disorder in QREM can be expressed in terms of strings of $\hat \sigma^z$ operators
\be
\sum_{\underline{\sigma}^z }^{2^L} E_{\underline{\sigma}^z}
 \ket{\underline{\sigma}^z} \bra{\underline{\sigma}^z} = \sum_k^L \sum_{i_1,..,i_k} J_{i_1,...,i_k}^{(k)} \hat\sigma_{i_1}^z\cdots \hat\sigma_{i_k}^z,
\ee
 with
$J_{i_1,...,i_k}^{(k)} = \frac{1}{2^L} \sum_{\underline{\sigma}_z} E_{\underline{\sigma}^z}  \sigma_{i_1}^z \cdots  \sigma_{i_k}^z$.
As compared with Hamiltonian $\hat H_{\text{MBL}}$, the QREM contains stronger interactions, and, for the system Eq.~\eqref{eq:Hamiltonian}, we expect
\be
W_c \le W_{\textrm{QREM}} \sim \sqrt{L}\log{L} < W_{AT} \sim L \ .
\ee
This imposes an even stronger upper bound on the possible scaling of the MBL transition with $L$, which is not consistent with~\cite{Lev19, Sels_2020}.

\subsection{Radial probability distribution $\Pi(x)$} \label{Sec:Radial}
\begin{figure}[h]
    \includegraphics[width=1.\linewidth]{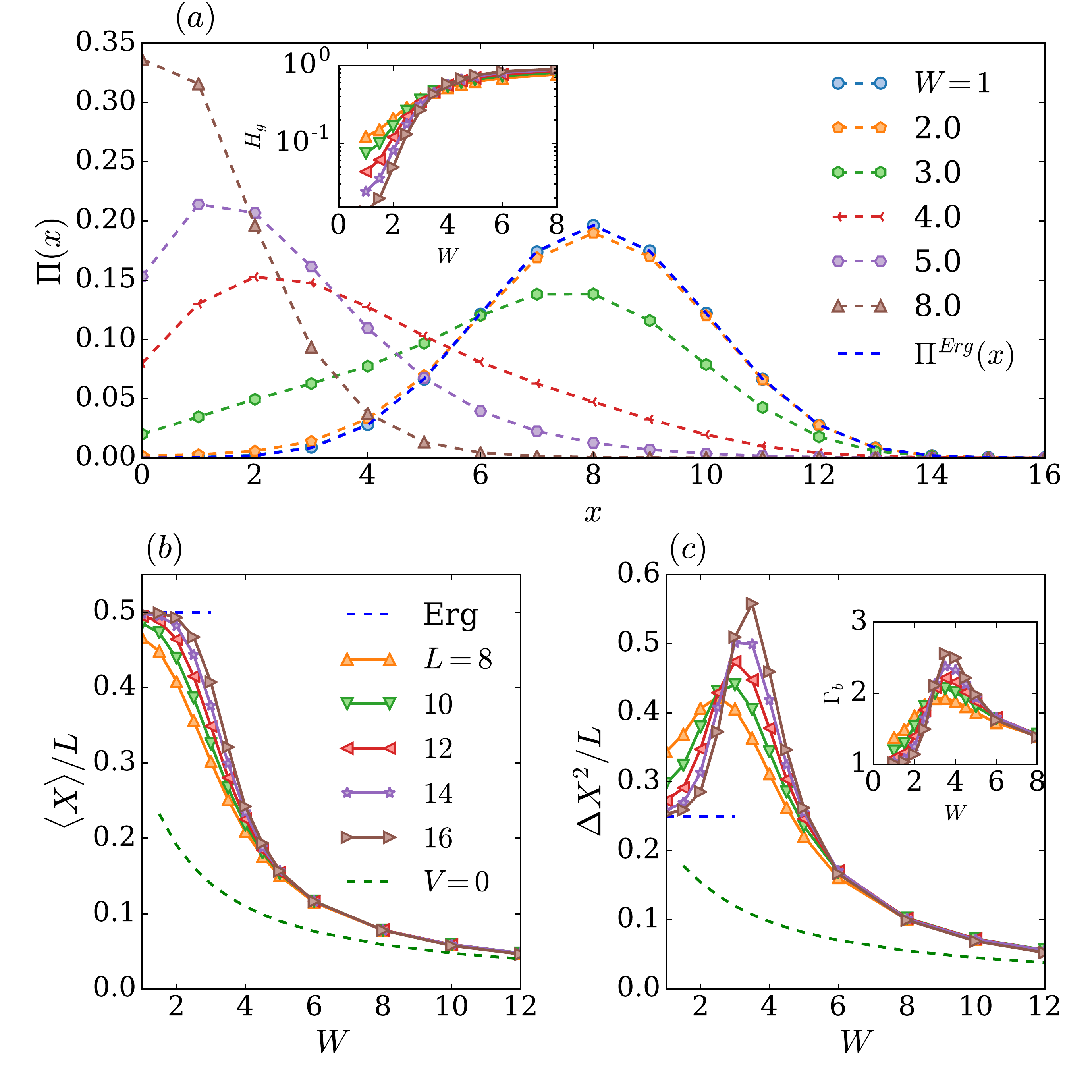}
    \caption{(a)~Radial probability distribution $\Pi(x)$ of the eigenstate coefficients in the Fock space, Eq.~\eqref{eq:Pi}, for a fixed system size $L=16$ and several disorder strengths $W$, ranging from the ergodic to the MBL phase (see legend). The blue dashed line show the ergodic prediction for $\Pi^{\text{Erg}}(x) = \frac{1}{2^L} \binom{L}{L/2}$.
    (inset) shows the Hellinger distance $H_g$, Eq.~\eqref{eq:Hellinger_dist}, between $\Pi(x)$ and the ergodic one $\Pi^{Erg}(x)$.
    (b)~Mean $\mean{X}/L$, Eq.~\eqref{eq:X}, and (c)~variance $\Delta X^2/L$, Eq.~\eqref{eq:X_2}, of $\Pi(x)$ as functions of $W$ for several $L$.
    Symbol and color code is shown in legend.
    The blue and green dashed lines correspond to the ergodic and the non-interacting ($V=0$) cases, respectively.
    The inset shows the deviation of $\Pi(x)$ from a binomial form~\eqref{eq:P(r)_p} via the ratio $\Gamma_b = L\Delta X^2/\mean{X}(L-\mean{X})$.
    }
    \label{fig:fig3}
\end{figure}

In this section, we show that the radial probability distribution $\Pi(x)$ in Eq.~\eqref{eq:Pi} of mid-spectrum eigenstate coefficients of $\hat H_{\text{MBL}}$ in the Fock space
gives more detailed information on the wave function's local structure in the Fock space compared to $\IPR_q$. In addition,
it can be related to the local integrals of motion of $\hat H_{\text{MBL}}$~\cite{serbyn2013local, huse2014phenomenology,ros2015integrals,Imbrie2016}.

We recall that at strong disorder, deep in the MBL phase, the eigenstates $\{\ket{E} \}$ of $\hat H_{\text{MBL}}$ are believed to be adiabatically connected to the non-interacting ones
$\{ \ket{\underline{\sigma}^z} \}$ through a sequence of quasi-local unitary operators,
 \begin{equation}
\ket{E} = \hat U \ket{\underline{\sigma}_0^z},
 \end{equation}
defining the integrals of motions $ \hat\tau_i^z =\hat U \hat \sigma^z_i \hat U^\dagger $ for which $[\hat H, \hat \tau_i^z] =0$.

Now, we use the above assumption to find the relation between the spread of the local integrals of motion $\{\hat \tau_i^z\}$ and
the moments of the probability distribution $\Pi(x)$. The Hamming distance between two Fock-states $\ket{\underline{\sigma}^z}$  and $\ket{\underline{\sigma}_0^z}$  is given by
 \begin{equation}
  d(\underline{\sigma}_0^z, \underline{\sigma}^z) = \sum_i \frac{(\sigma_i^z-\sigma_{0,i}^z)^2}{4},
 \end{equation}
where $\sigma_i^z=\bracket{\underline{\sigma}^z}{\hat \sigma_i^z}{\underline{\sigma}^z}$. The first moment of $\Pi(x,E)$, Eq.~\eqref{eq:X},
for a certain eigenstate at energy $E$ is given by,
\begin{equation}
 X(E) = \sum_x x \Pi(x,E) = \sum_{\sigma} d(\underline{\sigma}^z, \underline{\sigma}_0^z) \lrv{\bracket{\underline{\sigma}^z}{\hat U}{\underline{\sigma}_0^z}}^2,
\end{equation}
where we use $\ket{E} = U \ket{\underline{\sigma}_0^z}$. Thus,
\begin{multline}\label{eq:X(E)}
X(E) = \frac{L}{2} -  \frac{1}{2}\sum_i \sum_{\underline{\sigma}^z} \sigma_i^z \sigma_{0,i}^z \bracket{\underline{\sigma}_0^z}
{\hat U^\dagger}{\underline{\sigma}^z}  \bracket{\underline{\sigma}^z}{\hat U}{\underline{\sigma}_0^z} \\
= \frac{L}{2} - \frac{1}{2} \sum_i \bracket{E}{\hat \sigma_i^z  \hat \tau_i^z}{E} \ .
\end{multline}
Averaging over disorder and energies $E$, we obtain
\be\mean{X} = 
 \frac{L}{2} - \frac{1}{2} \mean{\sum_i  \bracket{E}{\hat \sigma_i^z  \hat \tau_i^z}{E} }
\ee
Similar calculations show that the variance in Eq.~\eqref{eq:X_2} of $\Pi(x, E)$, $\Delta X^2(E) =  \mean{X^2}(E) - \mean{X(E)}^2$, is given by
\begin{multline}\label{eq:d2X(E)}
 \Delta X^2(E) =
  \frac{1}{4} \sum_{i,j} \mean{\bracket{E}{\hat \sigma_i^z \hat \sigma_j^z \hat \tau_i^z \hat \tau_j^z }{E}} - \\
  -\frac{1}{4} \lrp{\sum_i \mean{\bracket{E}{ \hat \sigma_i^z \hat \tau_i^z }{E}}}^2.
\end{multline}
Thus, both first cumulants, $\mean{X}$ and $\Delta X^2$, provide the measures of the distance between the local integrals of motion $\{\hat \tau_i^z\}$ and the undressed operators $\{\hat \sigma_i^z\}$.
In particular, in the MBL phase we expect a perturbative expansion $\tau_i^z =   a_{i_1}^\alpha \hat \sigma_{i_1}^\alpha/\sqrt{\xi_{\text{loc}}} +  b_{i_1 i_2}^{\alpha\beta} \hat \sigma_{i_1}^\alpha \hat \sigma_{i_2}^\beta + \cdots$~\footnote{Repeated indices are summed.},
where $a_{i_1}^{\alpha} \sim e^{-\lrv{i-i_1}/\xi_{\text{loc}}}$, $b_{i_1 i_2}^{\alpha\beta} \sim e^{-\lrp{\lrv{i-i_1}+\lrv{i-i_2}}/\xi_{\text{loc}}}$ and Latin letters run over site indices and Greek ones are in $\{x,y,z\}$.
The coefficient $1/\sqrt{\xi_{\text{loc}}}$, with $a_{i}^\alpha \sim O(1)$, provides the normalization of the operator $\lrp{\tau_i^z}^2 = 1$.
Thus, $\mean{X}$ gives the direct estimate for the localization length
\be\label{eq:X_xi_loc}
\mean{X} \sim \frac{L}{2}\lrp{1-\sqrt{\frac{\xi_{\min}}{\xi_{\text{loc}}}} } \ ,
\ee
where $\xi_{\min}$ stands for some typical value of $(a_{i}^\alpha)^2$.

Having elucidated the relation between the integrals of motion and the radial probability distribution, we now present the numerical results for $\Pi(x)$, $\mean{X}$ and $\Delta X^2$.
Figure~\ref{fig:fig3}(a) shows $\Pi(x)$ for fixed system size $L=16$ and several disorder strengths.
As expected, at weak disorder, $\Pi(x)$ is centered in the middle of the chain, $\mean{X} = L/2$, with fluctuations $\Delta X^2 = L/4$ as shown
in Fig.~\ref{fig:fig3}(b-c).
In this case the shape of $\Pi(x)$ is close to the one $\Pi^{\text{Erg}}(x)$ of an ergodic system, Eq.~\eqref{eq:P(r)_p} with $p=\frac12$.
To quantify better the deviations of $\Pi(x)$ from $\Pi^{\text{Erg}}(x)$, we consider the Hellinger distance
\begin{equation}\label{eq:Hellinger_dist}
 H_g = \frac{1}{\sqrt{2}} \sqrt{\sum_x \left ( \sqrt{\Pi(x)} - \sqrt{\Pi^{\text{Erg}}(x)} \right )^2} \ ,
\end{equation}
which quantifies the distance between two probability distributions.
The inset in Fig.~\ref{fig:fig3}(a) shows $H_g$ as a function of $W$.
In agreement with the $\IPR_q$-analysis, at weak disorder $H_q$ tends to zero exponentially fast with $L$, since
the system is ergodic in the Hilbert space.
At strong disorder in the MBL phase, $\Pi(x)$ has a non-ergodic shape and it is skewed to the left with respect to its maximum,
meaning that $\mean{X} < L/2$, Fig.~\ref{fig:fig3}(a), and $H_g$ flows slowly with $L$ to its maximal value $H_g = 1$.

Both mean $\mean{X}/L$ and variance $\Delta X^2/L$ normalized by $L$ in Fig.~\ref{fig:fig3}(b-c), demonstrate similar behavior to the fractal dimension, $\mean{D_q}$, and the normalized entanglement entropy, $\mean{S}/L$ and its fluctuations shown in Fig.~\ref{fig:fig1}(a-b).
The quantity $\mean{X}/L$ decreases monotonically  with $W$ from its ergodic value (blue dashed line) toward the non-interacting one (green dashed curve).
At $W\lesssim W_c$, $\mean{X}/L$ flows with $L$ towards the ergodic value $\mean{X}/L = 1/2$, consistent with $\xi_{\text{loc}}^{-1}=0$ in  Eq.~\eqref{eq:X_xi_loc}. It shows the saturation with the system size in the MBL phase at the strictly positive value ($\xi_{\text{loc}}>\xi_{\min}$). This behavior is consistent with the finite jump in $\xi_{\text{loc}}^{-1}$ at the MBL transition.
The deviations of $\mean{X}/L$ and $\Delta X^2/L$ in the MBL phase from their non-interacting limits and the saturation of these deviations with the system size
are possibly related to the partial melting of the ergodic bubbles in the avalanche scenario~\cite{DeRoeck2018_MBL_Avalanche}.

Close to the transition $\Delta X^2/L$ exhibits a peak which diverges with the system size, $\Delta X^2 \sim L^{\alpha}$, with $\alpha >1$.
The latter evidences a rather broad profile of $\Pi(x)$ going beyond the binomial approximation, Eq.~\eqref{eq:P(r)_p}.
The inset to Fig.~\ref{fig:fig3}(c) emphasizes this by showing that the ratio $\Gamma_b = L\Delta X^2/\mean{X}(L-\mean{X})$ is strictly larger than its binomial unit value and demonstrates the same divergence with $L$ as the variance $\Delta X^2/L$.

To sum up in this section we have studied the radial probability distribution $\Pi(x)$, which is directly related to the local integrals of motions.
In particular, the mean $\mean{X}$ of $\Pi(x)$ can be used to define a localization length $\xi_{\text{loc}}$, Eq.~\eqref{eq:X_xi_loc}.
We have shown that at weak disorder $\mean{X} \rightarrow L/2$ corresponding to $\xi_{\text{loc}}\rightarrow \infty$, while
in the localized phase $0<\mean{X}/L<1/2$ giving finite $\xi_{\text{loc}}>\xi_{\min}$.
At the transition the fluctuations $\Delta X^2$ diverge which gives an evidence of a finite jump in $\mean{X}/L$
and therefore in $\xi_{\text{loc}}^{-1}$ in the thermodynamic limit $L\rightarrow \infty$.
The above finite jump in $\xi_{\text{loc}}^{-1}$ is consistent with the avalanche theory of many-body delocalization~\cite{DeRoeck2018_MBL_Avalanche} and therefore with the KT-type scaling of the transition~\cite{Goremykina2019-1,Goremykina2019-2}.

In the next section, we consider a simple toy model of dilute ergodic ``bubbles'', which explains the observed phenomenon of the absence of $D_q$ self-averaging,
and helps us to bridge the gap between Fock space structure of mid-spectrum eigenstates, the avalanche theory, and the recent studies of phenomenological renormalization groups of MBL transition.

\section{Dilute ergodic bubbles approximation}\label{Sec:Bubble}

In this section we discuss random block Hamiltonian consisting of non-interacting pieces which is able to reproduce the numerical results described in the previous section.
Ref.~\onlinecite{DeRoeck2018_MBL_Avalanche} proposes that non-perturbative effects, such as rare thermal inclusions, which are unavoidable due to entropic arguments, can destabilize the MBL phase provided their density exceeds a certain critical value.
As a consequence, an abrupt finite jump of the localization length is expected at the transition.
Moreover, RG-studies~\cite{Goremykina2019-1,Goremykina2019-2} have shown that this avalanche theory should lead to a KT-like scaling behavior and the probability distribution $P(\ell)$ of the size $\ell$ of the largest thermal bubble has a power-law fat tail leading to a diverging variance.

With the aim to keep the toy model as simple as possible and avoid introducing further ``fitting'' parameters, we consider
bimodal approximation of $\rho L$ active ergodic spins~\footnote{cf. the Supplementary material of~\cite{Mace_Laflorencie2019_XXZ}}
and $(1-\rho)L$ non-interacting (frozen) spins.
As $\Pi(x)$ is not sensitive to the phases of the eigenstate's coefficients, and
both ergodic and non-interacting limits of it are given by the binomial distribution in Eq.~\eqref{eq:P(r)_p},
we approximate the contributions of ergodic and frozen spins to the many-body wave function
as the non-interacting ones (Eq.~\eqref{eq:U_mat}), with
$\mean{\sin^2\tfrac{\theta_i}2}=\frac12$
for $\rho L$ active spins and
$\sin \theta_i = 1/\sqrt{1+h_i^2}$ for the remaining $(1-\rho) L$ frozen ones, as they are subject to strong fields
$h_i$. Due to self-averaging of the non-interacting contribution~\eqref{eq:dDq_2_self-av} we describe it by the mean value of $p$, Eq.~\eqref{eq:p}.
In this model, $\rho$ plays the rote of the density of ergodic bubbles.

Within the above approximation, one can easily find $\Pi(x)$ as follows.
At a spin-flip distance $x$ from the eigenstate maximum we take $k$ spin flips of {\it ergodic} type and $r-k$ of {\it frozen} type.
The number of such paths from the maximal configuration point in the Fock space is given by
\be
N_{k,x} = \lrp{\rho L\atop k}\lrp{(1-\rho) L \atop x - k}  \ ,
\ee
where the first combinatorial factor counts the number of possible selection of $k$ indistinguishable spin-flips of the ergodic type out of $\rho L$ available ones,
the second one counts the number of possible  $r-k$ indistinguishable spin-flips of the frozen type out of $(1-\rho) L$ available ones.
The corresponding wave function intensity averaged over $h_i$ will be
\be\label{eq:psi_erg_bubble}
\lrv{\psi_{k,x}}^2  = \lrp{\frac12}^{\rho L} p^{x-k} (1-p)^{(1-\rho) L - x + k} \ .
\ee

The total number of selections for any $k$ is given by
\be
N_x = \sum_{k=0}^{x} N_{k,x} = \lrp{ L \atop x}
\ .
\ee
As a result, one can obtain the probability distribution $\Pi(x,\rho)$ at the fixed ergodic spin density $\rho$
\begin{multline}
\label{eq:Pi_rho}
\Pi(x,\rho) = \sum_{k=0}^{x} N_{k,x}\lrv{\psi_{k,x}}^2  \\ =
\sum_{k=0}^{x} \Pi_{1/2}\lrp{\rho L,k} \Pi_{p}\lrp{(1-\rho) L,x-k}  
\end{multline}
which is simply given by  the convolution of two binomial distributions
\be
\Pi_{p'}(L',x') = \lrp {L'\atop x'} p'^{x'} (1-p')^{L'-x'} \ ,
\ee
one describing the ergodic bubble ($p'=1/2$) and the other the remaining frozen spins with $p'=p<1/2$.

Due by the simple nature of the model, the $\IPR_q$ is the product of the $\IPR_q$s for the $\rho L$ ergodic spins and the $(1-\rho)L$ frozen ones:
\begin{multline}
 \IPR_q = \sum_{x,k} N_{k,x} \lrv{\psi_{k,x}}^{2q} = \\=
 \lrp{\frac12}^{(q-1) \rho L}\lrb{p^q + (1-p)^q}^{(1-\rho)L},
\end{multline}
and thus the fractal dimensions, Eq.~\eqref{eq:D_q}, take the form
\be\label{eq:D_q_res}
D_q(\rho) = \rho + (1-\rho) \mean{D_q^0} \ ,
\ee
with  $\mean{D_q^0}$  the fractal dimension of the non-interacting frozen spins, which is a continuous function of $W$ with Eqs.~\eqref{eq:D_q>0},~\eqref{eq:D_q_non_int}.
In this approximation any jump in the fractal dimension $D_q(\rho)$ is directly related to the jump in the bubble density $\rho$.


By using the Gaussian approximation Eq.~\eqref{eq:X,dX_2_Gauss} for the binomial distribution Eq.~\eqref{eq:P(r)_p}, we can estimate  the first and second moment of $\Pi(x,\rho)$ in Eq.~\eqref{eq:Pi_rho}
\begin{align}
 \label{eq:r(rho)_av}
 \mean{X}_{\rho} & = \frac{L}{2} - \nu \ep L, \\
 \label{eq:r^2(rho)_av}
 \mean{X^2}_{\rho} &= \lrb{\frac{L}{2} - \nu\ep L}^2 + L\lrb{\frac14-\nu\ep^2}
\end{align}
and
\be\label{eq:rho_sigma_fe}
\Delta X^2_{\rho}=\mean{X^2}_{\rho}-\mean{X}_{\rho}^2=
L\lrb{\frac14-\nu\ep^2} \ ,
\ee
where $\ep = 1/2-p$ and $\nu = 1-\rho\leq 1$.
These results hold for fixed density $\rho$.
For $\rho = 0$ and $p$ given by~\eqref{eq:p}, they coincide with the
non-interacting results shown in Fig.~\ref{fig:fig3}~(b-c) as green dashed lines, which show plots of Eqs.~\eqref{eq:r(rho)_av} and Eq.~\eqref{eq:rho_sigma_fe} for these parameters.

We emphasize that
the ratio $\Gamma_b(\rho) = L\Delta X^2_\rho/\mean{X}_\rho(L-\mean{X}_\rho)$
is strictly smaller then its binomial value $\Gamma_b$ in case $0<\nu<1$, $\ep\ne 0$, and
\be
\frac{\mean{X}_{\rho}}{L}\lrp{L-\mean{X}_{\rho}} = L\lrb{\frac14-\nu^2\ep^2} \ ,
\ee
which contradicts the numerics in Fig.~\ref{fig:fig3}.
This observation underlines the importance of the fluctuations of $\rho$ and the corresponding disorder average.
Moreover, even the scaling of the maximum of $\Delta X^2 \sim L^{\alpha}$, $\alpha >1$, from the exact numerics with $L$ contradicts~\eqref{eq:rho_sigma_fe} with fixed~$\rho$.

In order to recover the observed behavior of $\Gamma_b>1$ in Fig.~\ref{fig:fig3}, it is crucial to take into account the non-trivial distribution $P(\rho)$ of the density of ergodic spins and perform the average over it.
After averaging Eqs.~\eqref{eq:D_q_res},~\eqref{eq:r(rho)_av}, and~\eqref{eq:r^2(rho)_av} and abbreviating
\be
\int_0^1 \rho P(\rho) d\rho = \bar \rho\equiv 1-\bar\nu, \quad
\int_0^1 \rho^2 P(\rho) d\rho  - (\bar \rho)^2 = \sigma_\rho^2,
\ee
one straightforwardly obtains
\begin{align}
\bar D_q & = 
1 - \bar\nu(1-D_q^0) \ , \\
\label{eq:X_res}
\mean{X} & = 
\frac{L}{2} - \bar\nu\ep L \\
\label{eq:dX_2_res}
\Delta X^2 & = 
\sigma_\rho^2\ep^2 L^2 + L\lrb{\frac14-\bar\nu \ep^2} \ .
\end{align}

The properties of the probability distribution $P(\rho)$ of the density of ergodic spins can be extracted from the works~\cite{DeRoeck2018_MBL_Avalanche, DeRoeck2017_MBL_RG,Goremykina2019-1,Goremykina2019-2}.
According to these works the distribution $P(\ell_i)$ of the length $\ell_i$ of a single ergodic bubble is (stretch)-exponential, $P(\ell_i)\sim e^{-c \ell_i^d}$,
or power-law, $P(\ell_i)\sim {\ell_i}^{-\alpha}$, $\alpha>2$ in the MBL phase depending on its type.
In the thermal phase, $P(\ell_i)$ is 
sharply concentrated on the length scale $L$ of the system size and tends to a delta-function for $ L \to \infty $, meaning the entire system is ergodic.
At the transition $P(\ell_i)$ is a power-law distributed $\sim \ell_i^{-\alpha_c}$ with the critical exponent $2\leq\alpha_c\leq 3$~\cite{DeRoeck2018_MBL_Avalanche},
in order to have finite mean $\mean{\ell}\sim L^0$ and diverging variance as $\mean{\ell^2}\sim L^{3-\alpha_c}$.

In our consideration we focus on the power-law probability distribution and determine the power $\alpha$ independently.
Due to entropic reasons,
the mean density $\mean{\rho}$ of the ergodic bubbles has to be finite even in the MBL phase.
This density can be approximated by the sum of individual bubbles of lengths $\ell_i$ normalized by the system size
\be
\label{eq:rho_tot}
\rho \approx \sum_{i=1}^K \frac{\ell_i}{L} \ ,
\ee
where $K$ is a cut-off ensuring finite total length of order $ L $ of the chain.
In the simplest approximation of independent identically distributed random $\ell_i$ one can apply the central limit theorem to $\rho$.

In the MBL phase where the mean and the variance of the length of a single ergodic bubble are finite, $\mean{\ell}\sim L^0$, $\mean{\ell^2}\sim L^0$, i.e., $\alpha>3$,
and the number of bubbles scales as $K\sim L$ due to the boundness of $\mean{\rho} = K\mean{\ell}/L\sim L^0$,
one obtains the Gaussian distribution of $\rho$ with the variance decaying as
\be\label{eq:sigma_rho_MBL}
\sigma_\rho^2(W>W_c) = \bar{\rho^2}-{\bar \rho}^2 \sim \frac{1}{L} \ ,
\ee
consistent with self-averaging. This also immediately confirms the self-averaging of $D_q$ via Eq.~\eqref{eq:D_q_res}.
In a ergodic phase, the variance $\sigma_\rho (W<W_c)$ is exponentially small in $L$ as $\mean{\rho}\to 1$.
At the MBL transition self-averaging of $\rho$ breaks down in this model only if $\mean{\ell^2}\to L^{3-\alpha_c}$  diverges in the thermodynamic limit, i.e.  $\alpha_c<3$.
Due to the finite mean density
\be
\mean{\rho} = \sum_{i=1}^K \frac{\mean{\ell_i}}{L} = \frac{K}{L} \max\lrp{L^0,L^{2-\alpha_c}} \sim L^0 \ ,
\ee
the number of ergodic bubbles scales as $K\sim \min\lrp{L,L^{\alpha_c-1}}$, which puts the lower bound on $\alpha_c>1$.
The corresponding scaling of the variance takes the form
\be\label{eq:sigma_rho_transition}
\sigma_\rho^2(W \simeq W_c) \sim \min\lrp{L^0,L^{2-\alpha_c}}, 1<\alpha_c<3 \ .
\ee
The condition of $\Delta X^2$ scaling faster than $L$ is in agreement with the observations in Fig.~\ref{fig:fig3}(c), and requires $\alpha_c < 3$, which ultimately
leads to the decay of the variance with $L$, but slower than $L^{-1}$.
As a consequence, both in the MBL phase and at the transition the most significant contribution to $\Delta X^2$ in~\eqref{eq:dX_2_res} is given by
the large fluctuations in $\rho$, cf.~Eqs.~\eqref{eq:sigma_rho_MBL},~\eqref{eq:sigma_rho_transition}.

The lack of self-averaging of $D_q$ in Eq.~\eqref{eq:D_q_res} implies the more restrictive condition $\alpha_c\leq 2$. This is consistent with the
finite mean $\mean{\ell}\sim L^0$ of a single bubble, $\alpha_c\geq 2$, only at the value $\alpha_c = 2$.

To summarize, in the ergodic phase $\rho = 1-\nu$ goes to unity exponentially $\sim e^{-\eta L}$ and its variance has to decay also at least exponentially and therefore
$\Delta X^2(W\ll W_c) \simeq L/4$, as expected in the ergodic case.
In the MBL phase $\mean{\rho}$ is finite and the variance decays as $1/L$ (cf.~\eqref{eq:sigma_rho_MBL}), which yields
a linear behavior of the mean $\mean{X}\sim L$ (cf.~Eq.~\eqref{eq:X_res}), and the variance $\Delta X^2 \sim L$ (cf.~Eq.~\eqref{eq:dX_2_res}).
At the transition, $W=W_c$, according to Eq.~\eqref{eq:sigma_rho_transition}, the variance of $\rho$ is large compared to $L^{-1}$ and, thus, it provides the main contribution to $\Delta X^2 \sim L^{4-\alpha_c}\gg L$, with $2\leq \alpha_c\leq 3$, i.e.
\begin{equation}
\Delta X^2/L \sim \begin{cases}
                 1/4 & W< W_c,\\
                 L^{3-\alpha_c} & W = W_c,\\
                 <1/4 & W>W_c.
               \end{cases}
\end{equation}
Finally, the constraint  $\alpha_c=2$ implies also that $D_q$ is not self-averaging at the transition.

Thus, we have shown that our numerical results can be explained by considering a simple toy model of dilute ergodic bubbles embedded into a sea of frozen clusters.
In particular, the divergence in $\Delta X^{2}/L$ requires to describe the probability distribution of the length of a typical ergodic bubble by a fat-tailed power-law distribution which has diverge fluctuations at the transition,
\begin{equation}
\Delta D_q^2 \sim \sigma^2_\rho \sim
                \begin{cases}
                 e^{-\eta L} & W< W_c,\\
                 L^{2-\alpha_c} & W = W_c,\\
                 L^{-1} & W>W_c.
               \end{cases}
\end{equation}

\section{Continuity of the transition for non-interacting systems} \label{Sec:continuous}
\begin{figure}
    \includegraphics[width=1.\linewidth]{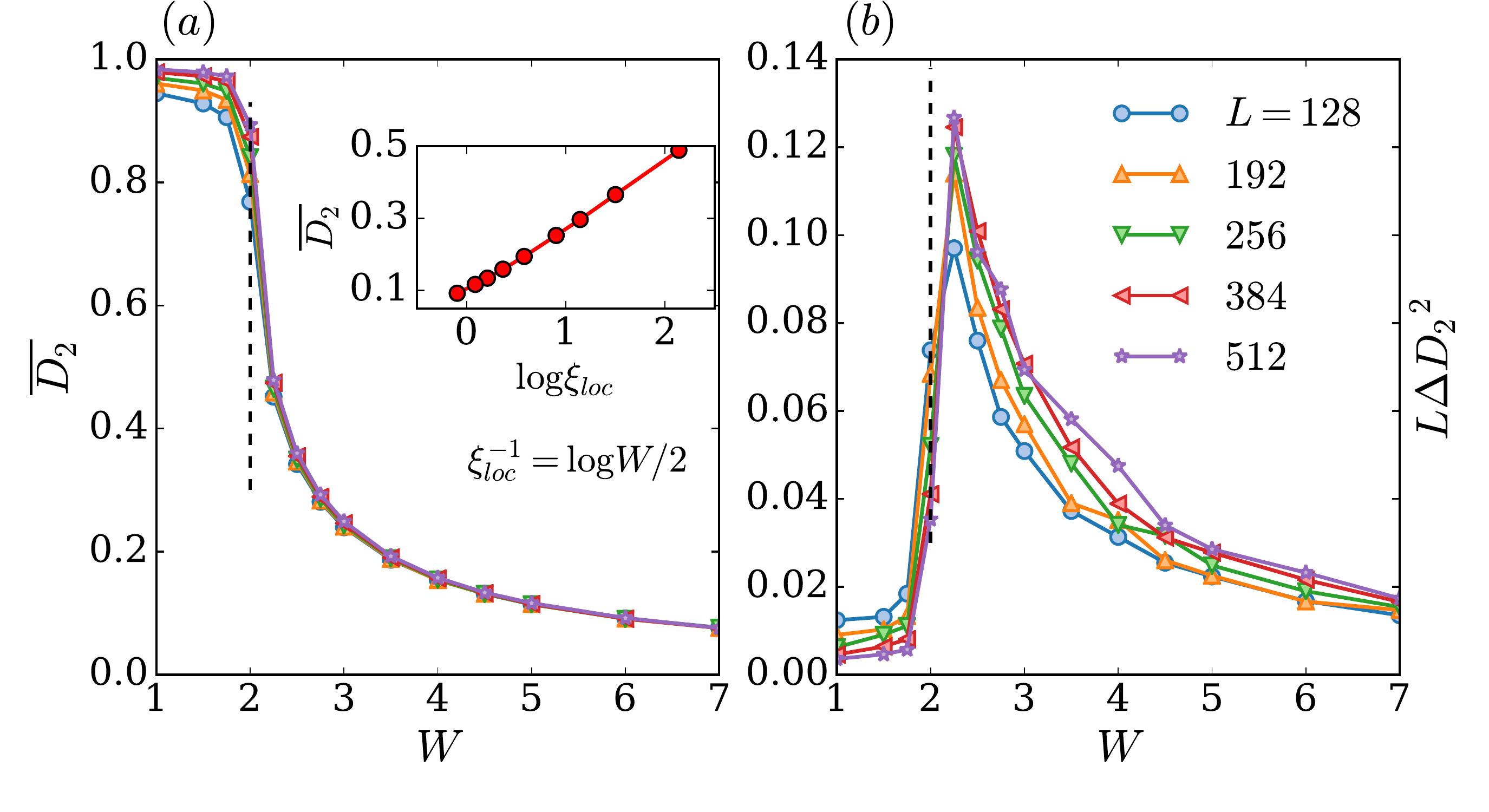}
    \caption{(a)~Fractal exponent $D_2$ and (b)~its variance $\Delta D_2^2$ for the non-interacting Aubry-Andr\'e model with the many-body fermionic half-filling for several $L$ versus the amplitude of the quasi-periodic potential $W$.
    The vertical black dashed line points to the critical point $W_c=2$.
    The $y$-axis of panel (b) has been rescaled to show that $D_2$ is self-averaging ($\Delta D_2^2 \sim 1/L$).
    The inset shows $\mean{D_2}$ for $W>2$ as a function of the localization length $\xi_{\text{loc}}^{-1} = \log{W/2}$.
    }
    \label{fig:fig4}
\end{figure}

In the previous sections, we have considered numerically the MBL transition from a Fock-space perspective. We have shown that it is could be characterized by a finite jump in the fractal dimensions $D_q$, which are not self-averaging at the critical point.
This scenario is consistent with the avalanche theory which predict a KT-type scaling at the transition and therefore a finite jump on the inverse localization length $\xi_{\text{loc}}^{-1}$.
For a contrast, we now investigate the case in which the system undergoes a delocalization-localization transition with continuous $\xi_{\text{loc}}^{-1}$, corresponding to the divergence of the localization length at the transition, $\lim_{W\rightarrow W_c} \xi_{\text{loc}} \rightarrow \infty$.
Concretely, we consider the non-interacting Aubrey-Andr\'e model
\be
\label{eq:H_AA}
\hat H_{AA} = -\frac{1}{2}\sum_i \hat c^\dagger_{i+1} \hat c_i + h.c. + W \sum_i h_i \hat c^\dagger_i \hat c_i,
\ee
where $\hat c^\dagger_i$ ($\hat c_i$) is the fermionic creation (annihilation) operator at site $i$,
$h_i= \cos{(2\pi \beta i + \delta)}$ is a quasi-periodic potential with $\beta = (\sqrt{5}+1)/2$,
and $\delta$ is a random phase uniformly distributed in the interval $[0,2\pi)$.
This single-particle model is known to have an Anderson transition at $W_c=2$ between extended ($W<2$) and localized ($W>2$) phases~\cite{AviJit2009}.
The single-particle localization length at transition diverges as $\xi_{\text{loc}}\sim 1/\log{W/2}$.
We consider non-interacting $\hat H_{AA}$ with the many-body fermionic filling and
focus on multifractal structure of the Slater-determinant many-body wave functions in the Fock space.
For this reason, we take the model on $L$ sites in the chain at half-filling $N=L/2$, where $N$ is the number of particles, maximizing the Hilbert space dimension $\binom{L}{N} \sim 2^{L}/\sqrt{L}$.
Fixing the basis of bare fermions $\ket{\underline{n}} = \prod_{n_i} (\hat c_{i}^\dagger)^{n_i} \ket{0}$ as the computational basis, with $n_i \in \{0,1\}$,
we rewrite the model $\hat H_{AA}$  in a similar form as Eq.~\eqref{eq:Hamiltonian_fock} and investigate the ergodic properties of its eigenstates in the Fock space.

Figure~\ref{fig:fig4}(a) shows the fractal dimension $\mean{D_2}$ for the eigenstates of $\hat H_{AA}$ constructed as a Slater determinant of taken at random $N$ single-particle eigenstates.
In the extended phase, $W<2$, the fractal dimension tends to unity, meaning that the typical eigenstate covers homogeneously the available Hilbert space.
Instead, in the single-particle localized phase, $W>2$, $\mean{D_2}$ converges to a strictly positive value smaller than one.
In fact, we expect that $IPR_2 \sim (\frac{1}{\xi_{\text{loc}}})^L$ similarly to a non-interacting spin chain, Eq.~\eqref{eq:D_q>0},
and as a consequence $D_2 \sim \log{\xi_{\text{loc}}}$ close to the transition.
Inset of Fig.~\ref{fig:fig4}(a) shows $\mean{D_2}$ as function $\log{\xi_{\text{loc}}}$ giving an evidence
of our prediction and therefore the many-body $D_2$ does not experience a jump across the transition.
Another important consequence is that $D_2$ is self-averaging as shown in Fig.~\ref{fig:fig4}(b), $\Delta D_2^2 \sim 1/L$, everywhere including the transition.
This self-averaging property should be compared to the one observed in the MBL model, for which the fractal dimensions are not self-averaging close to the MBL transition.

%
%

\section{Conclusion} \label{Sec:Conclusion}

In this work we have studied the MBL transition of a chain of interacting spins from a Fock-space point of view.
In addition to the standard diagnostic tools for ergodicity, such as fractal dimensions and entanglement entropy,
we consider the radial probability distribution of eigenstate coefficients with respect to
the Hamming distance in the Fock space from the wave function maximum.
We show that this radial probability distribution gives important insights about the integrals of motion of the problem and allows to extract the localization length
from the cumulants of the distribution.

Numerically we have found that both the fractal dimensions and the radial probability distribution have strong fluctuations at the critical point.
The fractal dimensions at the transition are not self-averaging and the probability distribution is extremely broad.
This divergence provides a rather strong evidence of the existence of a possible jump of the fractal dimensions as well as of the localization length across the MBL transition.

Inspired by recent studies of renormalization group and avalanche MBL theory, we explain our findings by introducing a simple spin toy model.
This model hosts a finite density of ergodic/thermal bubbles as well as frozen/localized spins.
The  MBL transition occurs by tuning the density of the ergodic bubbles.
At the transition the probability distribution of the bubble density is fat-tailed in agreement with the RG studies.
Using this simple model, we are able to explain our numerical findings, and thus to bridge the gap between recent studies of the nature of the MBL transition in the ``real space'' and in the Fock space perspectives.

As a result, we show that the MBL transition can be seen as a transition between ergodic states to non-ergodic extended states and we put an upper bound on the disorder scaling for a genuine Anderson localized regime with respect to the non-interacting case.

Finally, we provide an example of the many-body (non-interacting) model with a continuous localization transition
showing the self-averaging of fractal dimensions in the whole range of parameters.
This model confirms the conjectured relation between the non-self-averaging property of the fractal dimensions and their finite jump at the localization transition.

\begin{acknowledgments}
GDT acknowledges the hospitality of Max Planck Institute for the Physics of Complex Systems, Dresden, where part of the work was done.
IMK acknowledges the support of the Academy of Finland. This work was partially supported by the DFG under EXC-2111--390814868.
FP is funded by the European Research Council (ERC) under the European Unions Horizon 2020 research and innovation program (grant agreement No.  771537)  and under the DFG  TRR80.
\end{acknowledgments}

\bibliography{V_inf_bib}

\begin{thebibliography}{94}%
\makeatletter
\providecommand \@ifxundefined [1]{%
 \@ifx{#1\undefined}
}%
\providecommand \@ifnum [1]{%
 \ifnum #1\expandafter \@firstoftwo
 \else \expandafter \@secondoftwo
 \fi
}%
\providecommand \@ifx [1]{%
 \ifx #1\expandafter \@firstoftwo
 \else \expandafter \@secondoftwo
 \fi
}%
\providecommand \natexlab [1]{#1}%
\providecommand \enquote  [1]{``#1''}%
\providecommand \bibnamefont  [1]{#1}%
\providecommand \bibfnamefont [1]{#1}%
\providecommand \citenamefont [1]{#1}%
\providecommand \href@noop [0]{\@secondoftwo}%
\providecommand \href [0]{\begingroup \@sanitize@url \@href}%
\providecommand \@href[1]{\@@startlink{#1}\@@href}%
\providecommand \@@href[1]{\endgroup#1\@@endlink}%
\providecommand \@sanitize@url [0]{\catcode `\\12\catcode `\$12\catcode
  `\&12\catcode `\#12\catcode `\^12\catcode `\_12\catcode `\%12\relax}%
\providecommand \@@startlink[1]{}%
\providecommand \@@endlink[0]{}%
\providecommand \url  [0]{\begingroup\@sanitize@url \@url }%
\providecommand \@url [1]{\endgroup\@href {#1}{\urlprefix }}%
\providecommand \urlprefix  [0]{URL }%
\providecommand \Eprint [0]{\href }%
\providecommand \doibase [0]{http://dx.doi.org/}%
\providecommand \selectlanguage [0]{\@gobble}%
\providecommand \bibinfo  [0]{\@secondoftwo}%
\providecommand \bibfield  [0]{\@secondoftwo}%
\providecommand \translation [1]{[#1]}%
\providecommand \BibitemOpen [0]{}%
\providecommand \bibitemStop [0]{}%
\providecommand \bibitemNoStop [0]{.\EOS\space}%
\providecommand \EOS [0]{\spacefactor3000\relax}%
\providecommand \BibitemShut  [1]{\csname bibitem#1\endcsname}%
\let\auto@bib@innerbib\@empty
\bibitem [{\citenamefont {Deutsch}(1991)}]{Deutsch1991}%
  \BibitemOpen
  \bibfield  {author} {\bibinfo {author} {\bibfnamefont {J.~M.}\ \bibnamefont
  {Deutsch}},\ }\bibfield  {title} {\enquote {\bibinfo {title} {Quantum
  statistical mechanics in a closed system},}\ }\href {\doibase
  10.1103/PhysRevA.43.2046} {\bibfield  {journal} {\bibinfo  {journal} {Phys.
  Rev. A}\ }\textbf {\bibinfo {volume} {43}},\ \bibinfo {pages} {2046--2049}
  (\bibinfo {year} {1991})}\BibitemShut {NoStop}%
\bibitem [{\citenamefont {Srednicki}(1994)}]{Srednicki1994}%
  \BibitemOpen
  \bibfield  {author} {\bibinfo {author} {\bibfnamefont {Mark}\ \bibnamefont
  {Srednicki}},\ }\bibfield  {title} {\enquote {\bibinfo {title} {Chaos and
  quantum thermalization},}\ }\href {\doibase 10.1103/PhysRevE.50.888}
  {\bibfield  {journal} {\bibinfo  {journal} {Phys. Rev. E}\ }\textbf {\bibinfo
  {volume} {50}},\ \bibinfo {pages} {888--901} (\bibinfo {year}
  {1994})}\BibitemShut {NoStop}%
\bibitem [{\citenamefont {Srednicki}(1996)}]{Srednicki1996}%
  \BibitemOpen
  \bibfield  {author} {\bibinfo {author} {\bibfnamefont {Mark}\ \bibnamefont
  {Srednicki}},\ }\bibfield  {title} {\enquote {\bibinfo {title} {{Thermal
  fluctuations in quantized chaotic systems}},}\ }\href@noop {} {\bibfield
  {journal} {\bibinfo  {journal} {J. Phys. A: Mathematical and General}\
  }\textbf {\bibinfo {volume} {29}},\ \bibinfo {pages} {L75} (\bibinfo {year}
  {1996})}\BibitemShut {NoStop}%
\bibitem [{\citenamefont {{Rigol Marcos}}\ \emph {et~al.}(2008)\citenamefont
  {{Rigol Marcos}}, \citenamefont {{Dunjko Vanja}},\ and\ \citenamefont
  {{Olshanii Maxim}}}]{rigol2008thermalization}%
  \BibitemOpen
  \bibfield  {author} {\bibinfo {author} {\bibnamefont {{Rigol Marcos}}},
  \bibinfo {author} {\bibnamefont {{Dunjko Vanja}}}, \ and\ \bibinfo {author}
  {\bibnamefont {{Olshanii Maxim}}},\ }\bibfield  {title} {\enquote {\bibinfo
  {title} {{Thermalization and its mechanism for generic isolated quantum
  systems}},}\ }\href@noop {} {\bibfield  {journal} {\bibinfo  {journal}
  {Nature}\ }\textbf {\bibinfo {volume} {452}},\ \bibinfo {pages} {854}
  (\bibinfo {year} {2008})}\BibitemShut {NoStop}%
\bibitem [{\citenamefont {Polkovnikov}\ \emph {et~al.}(2011)\citenamefont
  {Polkovnikov}, \citenamefont {Sengupta}, \citenamefont {Silva},\ and\
  \citenamefont {Vengalattore}}]{Polkon_2011}%
  \BibitemOpen
  \bibfield  {author} {\bibinfo {author} {\bibfnamefont {Anatoli}\ \bibnamefont
  {Polkovnikov}}, \bibinfo {author} {\bibfnamefont {Krishnendu}\ \bibnamefont
  {Sengupta}}, \bibinfo {author} {\bibfnamefont {Alessandro}\ \bibnamefont
  {Silva}}, \ and\ \bibinfo {author} {\bibfnamefont {Mukund}\ \bibnamefont
  {Vengalattore}},\ }\bibfield  {title} {\enquote {\bibinfo {title}
  {Colloquium: Nonequilibrium dynamics of closed interacting quantum
  systems},}\ }\href {\doibase 10.1103/RevModPhys.83.863} {\bibfield  {journal}
  {\bibinfo  {journal} {Rev. Mod. Phys.}\ }\textbf {\bibinfo {volume} {83}},\
  \bibinfo {pages} {863--883} (\bibinfo {year} {2011})}\BibitemShut {NoStop}%
\bibitem [{\citenamefont {D'Alessio}\ \emph {et~al.}(2016)\citenamefont
  {D'Alessio}, \citenamefont {Kafri}, \citenamefont {Polkovnikov},\ and\
  \citenamefont {Rigol}}]{DAlessio2016ETH}%
  \BibitemOpen
  \bibfield  {author} {\bibinfo {author} {\bibfnamefont {Luca}\ \bibnamefont
  {D'Alessio}}, \bibinfo {author} {\bibfnamefont {Yariv}\ \bibnamefont
  {Kafri}}, \bibinfo {author} {\bibfnamefont {Anatoli}\ \bibnamefont
  {Polkovnikov}}, \ and\ \bibinfo {author} {\bibfnamefont {Marcos}\
  \bibnamefont {Rigol}},\ }\bibfield  {title} {\enquote {\bibinfo {title} {From
  quantum chaos and eigenstate thermalization to statistical mechanics and
  thermodynamics},}\ }\href {\doibase 10.1080/00018732.2016.1198134} {\bibfield
   {journal} {\bibinfo  {journal} {Advances in Physics}\ }\textbf {\bibinfo
  {volume} {65}},\ \bibinfo {pages} {239--362} (\bibinfo {year}
  {2016})}\BibitemShut {NoStop}%
\bibitem [{\citenamefont {Basko}\ \emph {et~al.}(2006)\citenamefont {Basko},
  \citenamefont {Aleiner},\ and\ \citenamefont {Altshuler}}]{Basko06}%
  \BibitemOpen
  \bibfield  {author} {\bibinfo {author} {\bibfnamefont {D.M.}\ \bibnamefont
  {Basko}}, \bibinfo {author} {\bibfnamefont {I.L.}\ \bibnamefont {Aleiner}}, \
  and\ \bibinfo {author} {\bibfnamefont {B.L.}\ \bibnamefont {Altshuler}},\
  }\bibfield  {title} {\enquote {\bibinfo {title} {Metal-insulator transition
  in a weakly interacting many-electron system with localized single-particle
  states},}\ }\href {\doibase https://doi.org/10.1016/j.aop.2005.11.014}
  {\bibfield  {journal} {\bibinfo  {journal} {Annals of Physics}\ }\textbf
  {\bibinfo {volume} {321}},\ \bibinfo {pages} {1126 -- 1205} (\bibinfo {year}
  {2006})}\BibitemShut {NoStop}%
\bibitem [{\citenamefont {Gornyi}\ \emph {et~al.}(2005)\citenamefont {Gornyi},
  \citenamefont {Mirlin},\ and\ \citenamefont
  {Polyakov}}]{gornyi2005interacting}%
  \BibitemOpen
  \bibfield  {author} {\bibinfo {author} {\bibfnamefont {IV}~\bibnamefont
  {Gornyi}}, \bibinfo {author} {\bibfnamefont {AD}~\bibnamefont {Mirlin}}, \
  and\ \bibinfo {author} {\bibfnamefont {DG}~\bibnamefont {Polyakov}},\
  }\bibfield  {title} {\enquote {\bibinfo {title} {{Interacting electrons in
  disordered wires: {A}nderson localization and low-{T} transport}},}\
  }\href@noop {} {\bibfield  {journal} {\bibinfo  {journal} {Phys. Rev. Lett.}\
  }\textbf {\bibinfo {volume} {95}},\ \bibinfo {pages} {206603} (\bibinfo
  {year} {2005})}\BibitemShut {NoStop}%
\bibitem [{\citenamefont {Pal}\ and\ \citenamefont {Huse}(2010)}]{Pal2010}%
  \BibitemOpen
  \bibfield  {author} {\bibinfo {author} {\bibfnamefont {Arijeet}\ \bibnamefont
  {Pal}}\ and\ \bibinfo {author} {\bibfnamefont {David~A.}\ \bibnamefont
  {Huse}},\ }\bibfield  {title} {\enquote {\bibinfo {title} {Many-body
  localization phase transition},}\ }\href {\doibase
  10.1103/PhysRevB.82.174411} {\bibfield  {journal} {\bibinfo  {journal} {Phys.
  Rev. B}\ }\textbf {\bibinfo {volume} {82}},\ \bibinfo {pages} {174411}
  (\bibinfo {year} {2010})}\BibitemShut {NoStop}%
\bibitem [{\citenamefont {Oganesyan}\ and\ \citenamefont
  {Huse}(2007)}]{oganesyan2007localization}%
  \BibitemOpen
  \bibfield  {author} {\bibinfo {author} {\bibfnamefont {Vadim}\ \bibnamefont
  {Oganesyan}}\ and\ \bibinfo {author} {\bibfnamefont {David~A.}\ \bibnamefont
  {Huse}},\ }\bibfield  {title} {\enquote {\bibinfo {title} {Localization of
  interacting fermions at high temperature},}\ }\href {\doibase
  10.1103/PhysRevB.75.155111} {\bibfield  {journal} {\bibinfo  {journal} {Phys.
  Rev. B}\ }\textbf {\bibinfo {volume} {75}},\ \bibinfo {pages} {155111}
  (\bibinfo {year} {2007})}\BibitemShut {NoStop}%
\bibitem [{\citenamefont {Alet}\ and\ \citenamefont
  {Laflorencie}(2018)}]{ALET2018498}%
  \BibitemOpen
  \bibfield  {author} {\bibinfo {author} {\bibfnamefont {Fabien}\ \bibnamefont
  {Alet}}\ and\ \bibinfo {author} {\bibfnamefont {Nicolas}\ \bibnamefont
  {Laflorencie}},\ }\bibfield  {title} {\enquote {\bibinfo {title} {Many-body
  localization: An introduction and selected topics},}\ }\href {\doibase
  https://doi.org/10.1016/j.crhy.2018.03.003} {\bibfield  {journal} {\bibinfo
  {journal} {Comptes Rendus Physique}\ }\textbf {\bibinfo {volume} {19}},\
  \bibinfo {pages} {498 -- 525} (\bibinfo {year} {2018})},\ \bibinfo {note}
  {quantum simulation / Simulation quantique}\BibitemShut {NoStop}%
\bibitem [{\citenamefont {Nandkishore}\ and\ \citenamefont
  {Huse}(2015)}]{huse2015review}%
  \BibitemOpen
  \bibfield  {author} {\bibinfo {author} {\bibfnamefont {Rahul}\ \bibnamefont
  {Nandkishore}}\ and\ \bibinfo {author} {\bibfnamefont {David~A.}\
  \bibnamefont {Huse}},\ }\bibfield  {title} {\enquote {\bibinfo {title}
  {{Many-Body Localization and Thermalization in Quantum Statistical
  Mechanics}},}\ }\href@noop {} {\bibfield  {journal} {\bibinfo  {journal}
  {Annual Review of Condensed Matter Physics}\ }\textbf {\bibinfo {volume}
  {6}},\ \bibinfo {pages} {15--38} (\bibinfo {year} {2015})}\BibitemShut
  {NoStop}%
\bibitem [{\citenamefont {Abanin}\ \emph {et~al.}(2019)\citenamefont {Abanin},
  \citenamefont {Altman}, \citenamefont {Bloch},\ and\ \citenamefont
  {Serbyn}}]{CollAba}%
  \BibitemOpen
  \bibfield  {author} {\bibinfo {author} {\bibfnamefont {Dmitry~A.}\
  \bibnamefont {Abanin}}, \bibinfo {author} {\bibfnamefont {Ehud}\ \bibnamefont
  {Altman}}, \bibinfo {author} {\bibfnamefont {Immanuel}\ \bibnamefont
  {Bloch}}, \ and\ \bibinfo {author} {\bibfnamefont {Maksym}\ \bibnamefont
  {Serbyn}},\ }\bibfield  {title} {\enquote {\bibinfo {title} {Colloquium:
  Many-body localization, thermalization, and entanglement},}\ }\href {\doibase
  10.1103/RevModPhys.91.021001} {\bibfield  {journal} {\bibinfo  {journal}
  {Rev. Mod. Phys.}\ }\textbf {\bibinfo {volume} {91}},\ \bibinfo {pages}
  {021001} (\bibinfo {year} {2019})}\BibitemShut {NoStop}%
\bibitem [{\citenamefont {Serbyn}\ \emph {et~al.}(2013)\citenamefont {Serbyn},
  \citenamefont {Papi{\'c}},\ and\ \citenamefont {Abanin}}]{serbyn2013local}%
  \BibitemOpen
  \bibfield  {author} {\bibinfo {author} {\bibfnamefont {Maksym}\ \bibnamefont
  {Serbyn}}, \bibinfo {author} {\bibfnamefont {Z}~\bibnamefont {Papi{\'c}}}, \
  and\ \bibinfo {author} {\bibfnamefont {Dmitry~A}\ \bibnamefont {Abanin}},\
  }\bibfield  {title} {\enquote {\bibinfo {title} {Local conservation laws and
  the structure of the many-body localized states},}\ }\href
  {http://dx.doi.org/10.1103/PhysRevLett.111.127201} {\bibfield  {journal}
  {\bibinfo  {journal} {Phys. Rev. Lett.}\ }\textbf {\bibinfo {volume} {111}},\
  \bibinfo {pages} {127201} (\bibinfo {year} {2013})}\BibitemShut {NoStop}%
\bibitem [{\citenamefont {Huse}\ \emph {et~al.}(2014)\citenamefont {Huse},
  \citenamefont {Nandkishore},\ and\ \citenamefont
  {Oganesyan}}]{huse2014phenomenology}%
  \BibitemOpen
  \bibfield  {author} {\bibinfo {author} {\bibfnamefont {David~A}\ \bibnamefont
  {Huse}}, \bibinfo {author} {\bibfnamefont {Rahul}\ \bibnamefont
  {Nandkishore}}, \ and\ \bibinfo {author} {\bibfnamefont {Vadim}\ \bibnamefont
  {Oganesyan}},\ }\bibfield  {title} {\enquote {\bibinfo {title}
  {{Phenomenology of certain many-body-localized systems}},}\ }\href
  {https://journals.aps.org/prb/pdf/10.1103/PhysRevB.90.174202} {\bibfield
  {journal} {\bibinfo  {journal} {Phys. Rev. B}\ }\textbf {\bibinfo {volume}
  {90}},\ \bibinfo {pages} {174202} (\bibinfo {year} {2014})}\BibitemShut
  {NoStop}%
\bibitem [{\citenamefont {Ros}\ \emph {et~al.}(2015)\citenamefont {Ros},
  \citenamefont {M{\"u}ller},\ and\ \citenamefont
  {Scardicchio}}]{ros2015integrals}%
  \BibitemOpen
  \bibfield  {author} {\bibinfo {author} {\bibfnamefont {V}~\bibnamefont
  {Ros}}, \bibinfo {author} {\bibfnamefont {M}~\bibnamefont {M{\"u}ller}}, \
  and\ \bibinfo {author} {\bibfnamefont {A}~\bibnamefont {Scardicchio}},\
  }\bibfield  {title} {\enquote {\bibinfo {title} {{Integrals of motion in the
  many-body localized phase}},}\ }\href
  {http://www.sciencedirect.com/science/article/pii/S0550321314003836}
  {\bibfield  {journal} {\bibinfo  {journal} {Nucl. Phys. B}\ }\textbf
  {\bibinfo {volume} {891}},\ \bibinfo {pages} {420--465} (\bibinfo {year}
  {2015})}\BibitemShut {NoStop}%
\bibitem [{\citenamefont {Luitz}\ \emph {et~al.}(2015)\citenamefont {Luitz},
  \citenamefont {Laflorencie},\ and\ \citenamefont {Alet}}]{Luitz15}%
  \BibitemOpen
  \bibfield  {author} {\bibinfo {author} {\bibfnamefont {David~J.}\
  \bibnamefont {Luitz}}, \bibinfo {author} {\bibfnamefont {Nicolas}\
  \bibnamefont {Laflorencie}}, \ and\ \bibinfo {author} {\bibfnamefont
  {Fabien}\ \bibnamefont {Alet}},\ }\bibfield  {title} {\enquote {\bibinfo
  {title} {Many-body localization edge in the random-field {H}eisenberg
  chain},}\ }\href {\doibase 10.1103/PhysRevB.91.081103} {\bibfield  {journal}
  {\bibinfo  {journal} {Phys. Rev. B}\ }\textbf {\bibinfo {volume} {91}},\
  \bibinfo {pages} {081103} (\bibinfo {year} {2015})}\BibitemShut {NoStop}%
\bibitem [{\citenamefont {Canovi}\ \emph {et~al.}(2011)\citenamefont {Canovi},
  \citenamefont {Rossini}, \citenamefont {Fazio}, \citenamefont {Santoro},\
  and\ \citenamefont {Silva}}]{Canovi11}%
  \BibitemOpen
  \bibfield  {author} {\bibinfo {author} {\bibfnamefont {Elena}\ \bibnamefont
  {Canovi}}, \bibinfo {author} {\bibfnamefont {Davide}\ \bibnamefont
  {Rossini}}, \bibinfo {author} {\bibfnamefont {Rosario}\ \bibnamefont
  {Fazio}}, \bibinfo {author} {\bibfnamefont {Giuseppe~E.}\ \bibnamefont
  {Santoro}}, \ and\ \bibinfo {author} {\bibfnamefont {Alessandro}\
  \bibnamefont {Silva}},\ }\bibfield  {title} {\enquote {\bibinfo {title}
  {Quantum quenches, thermalization, and many-body localization},}\ }\href
  {\doibase 10.1103/PhysRevB.83.094431} {\bibfield  {journal} {\bibinfo
  {journal} {Phys. Rev. B}\ }\textbf {\bibinfo {volume} {83}},\ \bibinfo
  {pages} {094431} (\bibinfo {year} {2011})}\BibitemShut {NoStop}%
\bibitem [{\citenamefont {Kj\"all}\ \emph {et~al.}(2014)\citenamefont
  {Kj\"all}, \citenamefont {Bardarson},\ and\ \citenamefont
  {Pollmann}}]{Kill14}%
  \BibitemOpen
  \bibfield  {author} {\bibinfo {author} {\bibfnamefont {Jonas~A.}\
  \bibnamefont {Kj\"all}}, \bibinfo {author} {\bibfnamefont {Jens~H.}\
  \bibnamefont {Bardarson}}, \ and\ \bibinfo {author} {\bibfnamefont {Frank}\
  \bibnamefont {Pollmann}},\ }\bibfield  {title} {\enquote {\bibinfo {title}
  {Many-body localization in a disordered quantum {I}sing chain},}\ }\href
  {\doibase 10.1103/PhysRevLett.113.107204} {\bibfield  {journal} {\bibinfo
  {journal} {Phys. Rev. Lett.}\ }\textbf {\bibinfo {volume} {113}},\ \bibinfo
  {pages} {107204} (\bibinfo {year} {2014})}\BibitemShut {NoStop}%
\bibitem [{\citenamefont {De~Tomasi}\ \emph {et~al.}(2017)\citenamefont
  {De~Tomasi}, \citenamefont {Bera}, \citenamefont {Bardarson},\ and\
  \citenamefont {Pollmann}}]{Giu17}%
  \BibitemOpen
  \bibfield  {author} {\bibinfo {author} {\bibfnamefont {Giuseppe}\
  \bibnamefont {De~Tomasi}}, \bibinfo {author} {\bibfnamefont {Soumya}\
  \bibnamefont {Bera}}, \bibinfo {author} {\bibfnamefont {Jens~H.}\
  \bibnamefont {Bardarson}}, \ and\ \bibinfo {author} {\bibfnamefont {Frank}\
  \bibnamefont {Pollmann}},\ }\bibfield  {title} {\enquote {\bibinfo {title}
  {Quantum mutual information as a probe for many-body localization},}\ }\href
  {\doibase 10.1103/PhysRevLett.118.016804} {\bibfield  {journal} {\bibinfo
  {journal} {Phys. Rev. Lett.}\ }\textbf {\bibinfo {volume} {118}},\ \bibinfo
  {pages} {016804} (\bibinfo {year} {2017})}\BibitemShut {NoStop}%
\bibitem [{\citenamefont {Vardhan}\ \emph {et~al.}(2017)\citenamefont
  {Vardhan}, \citenamefont {De~Tomasi}, \citenamefont {Heyl}, \citenamefont
  {Heller},\ and\ \citenamefont {Pollmann}}]{Var17}%
  \BibitemOpen
  \bibfield  {author} {\bibinfo {author} {\bibfnamefont {Shreya}\ \bibnamefont
  {Vardhan}}, \bibinfo {author} {\bibfnamefont {Giuseppe}\ \bibnamefont
  {De~Tomasi}}, \bibinfo {author} {\bibfnamefont {Markus}\ \bibnamefont
  {Heyl}}, \bibinfo {author} {\bibfnamefont {Eric~J.}\ \bibnamefont {Heller}},
  \ and\ \bibinfo {author} {\bibfnamefont {Frank}\ \bibnamefont {Pollmann}},\
  }\bibfield  {title} {\enquote {\bibinfo {title} {Characterizing time
  irreversibility in disordered fermionic systems by the effect of local
  perturbations},}\ }\href {\doibase 10.1103/PhysRevLett.119.016802} {\bibfield
   {journal} {\bibinfo  {journal} {Phys. Rev. Lett.}\ }\textbf {\bibinfo
  {volume} {119}},\ \bibinfo {pages} {016802} (\bibinfo {year}
  {2017})}\BibitemShut {NoStop}%
\bibitem [{\citenamefont {Gornyi}\ \emph {et~al.}(2016)\citenamefont {Gornyi},
  \citenamefont {Mirlin},\ and\ \citenamefont {Polyakov}}]{Mirlin2016dot}%
  \BibitemOpen
  \bibfield  {author} {\bibinfo {author} {\bibfnamefont {I.~V.}\ \bibnamefont
  {Gornyi}}, \bibinfo {author} {\bibfnamefont {A.~D.}\ \bibnamefont {Mirlin}},
  \ and\ \bibinfo {author} {\bibfnamefont {D.~G.}\ \bibnamefont {Polyakov}},\
  }\bibfield  {title} {\enquote {\bibinfo {title} {{Many-body delocalization
  transition and relaxation in a quantum dot}},}\ }\href@noop {} {\bibfield
  {journal} {\bibinfo  {journal} {Phys. Rev. B}\ }\textbf {\bibinfo {volume}
  {93}},\ \bibinfo {pages} {125419} (\bibinfo {year} {2016})}\BibitemShut
  {NoStop}%
\bibitem [{\citenamefont {Khemani}\ \emph {et~al.}(2017)\citenamefont
  {Khemani}, \citenamefont {Lim}, \citenamefont {Sheng},\ and\ \citenamefont
  {Huse}}]{Vedi17}%
  \BibitemOpen
  \bibfield  {author} {\bibinfo {author} {\bibfnamefont {Vedika}\ \bibnamefont
  {Khemani}}, \bibinfo {author} {\bibfnamefont {S.~P.}\ \bibnamefont {Lim}},
  \bibinfo {author} {\bibfnamefont {D.~N.}\ \bibnamefont {Sheng}}, \ and\
  \bibinfo {author} {\bibfnamefont {David~A.}\ \bibnamefont {Huse}},\
  }\bibfield  {title} {\enquote {\bibinfo {title} {Critical properties of the
  many-body localization transition},}\ }\href {\doibase
  10.1103/PhysRevX.7.021013} {\bibfield  {journal} {\bibinfo  {journal} {Phys.
  Rev. X}\ }\textbf {\bibinfo {volume} {7}},\ \bibinfo {pages} {021013}
  (\bibinfo {year} {2017})}\BibitemShut {NoStop}%
\bibitem [{\citenamefont {Vosk}\ \emph {et~al.}(2015)\citenamefont {Vosk},
  \citenamefont {Huse},\ and\ \citenamefont {Altman}}]{Vosk15}%
  \BibitemOpen
  \bibfield  {author} {\bibinfo {author} {\bibfnamefont {Ronen}\ \bibnamefont
  {Vosk}}, \bibinfo {author} {\bibfnamefont {David~A.}\ \bibnamefont {Huse}}, \
  and\ \bibinfo {author} {\bibfnamefont {Ehud}\ \bibnamefont {Altman}},\
  }\bibfield  {title} {\enquote {\bibinfo {title} {Theory of the many-body
  localization transition in one-dimensional systems},}\ }\href {\doibase
  10.1103/PhysRevX.5.031032} {\bibfield  {journal} {\bibinfo  {journal} {Phys.
  Rev. X}\ }\textbf {\bibinfo {volume} {5}},\ \bibinfo {pages} {031032}
  (\bibinfo {year} {2015})}\BibitemShut {NoStop}%
\bibitem [{\citenamefont {Bera}\ \emph {et~al.}(2015)\citenamefont {Bera},
  \citenamefont {Schomerus}, \citenamefont {Heidrich-Meisner},\ and\
  \citenamefont {Bardarson}}]{Bera15}%
  \BibitemOpen
  \bibfield  {author} {\bibinfo {author} {\bibfnamefont {Soumya}\ \bibnamefont
  {Bera}}, \bibinfo {author} {\bibfnamefont {Henning}\ \bibnamefont
  {Schomerus}}, \bibinfo {author} {\bibfnamefont {Fabian}\ \bibnamefont
  {Heidrich-Meisner}}, \ and\ \bibinfo {author} {\bibfnamefont {Jens~H.}\
  \bibnamefont {Bardarson}},\ }\bibfield  {title} {\enquote {\bibinfo {title}
  {Many-body localization characterized from a one-particle perspective},}\
  }\href {\doibase 10.1103/PhysRevLett.115.046603} {\bibfield  {journal}
  {\bibinfo  {journal} {Phys. Rev. Lett.}\ }\textbf {\bibinfo {volume} {115}},\
  \bibinfo {pages} {046603} (\bibinfo {year} {2015})}\BibitemShut {NoStop}%
\bibitem [{\citenamefont {{Chandran}}\ \emph {et~al.}(2015)\citenamefont
  {{Chandran}}, \citenamefont {{Laumann}},\ and\ \citenamefont
  {{Oganesyan}}}]{Chandran_2015_bounds}%
  \BibitemOpen
  \bibfield  {author} {\bibinfo {author} {\bibfnamefont {A.}~\bibnamefont
  {{Chandran}}}, \bibinfo {author} {\bibfnamefont {C.~R.}\ \bibnamefont
  {{Laumann}}}, \ and\ \bibinfo {author} {\bibfnamefont {V.}~\bibnamefont
  {{Oganesyan}}},\ }\bibfield  {title} {\enquote {\bibinfo {title} {{Finite
  size scaling bounds on many-body localized phase transitions}},}\ }\href@noop
  {} {\bibfield  {journal} {\bibinfo  {journal} {arXiv e-prints}\ ,\ \bibinfo
  {eid} {arXiv:1509.04285}} (\bibinfo {year} {2015})},\ \Eprint
  {http://arxiv.org/abs/1509.04285} {arXiv:1509.04285 [cond-mat.dis-nn]}
  \BibitemShut {NoStop}%
\bibitem [{\citenamefont {Thiery}\ \emph {et~al.}(2017)\citenamefont {Thiery},
  \citenamefont {M{\"u}ller},\ and\ \citenamefont
  {De~Roeck}}]{DeRoeck2017_MBL_RG}%
  \BibitemOpen
  \bibfield  {author} {\bibinfo {author} {\bibfnamefont {Thimoth{\'e}e}\
  \bibnamefont {Thiery}}, \bibinfo {author} {\bibfnamefont {Markus}\
  \bibnamefont {M{\"u}ller}}, \ and\ \bibinfo {author} {\bibfnamefont
  {Wojciech}\ \bibnamefont {De~Roeck}},\ }\href@noop {} {\enquote {\bibinfo
  {title} {A microscopically motivated renormalization scheme for the
  {M}{B}{L}/{E}{T}{H} transition},}\ } (\bibinfo {year} {2017}),\ \Eprint
  {http://arxiv.org/abs/1711.09880} {arXiv:1711.09880} \BibitemShut {NoStop}%
\bibitem [{\citenamefont {Thiery}\ \emph {et~al.}(2018)\citenamefont {Thiery},
  \citenamefont {Huveneers}, \citenamefont {M\"uller},\ and\ \citenamefont
  {De~Roeck}}]{DeRoeck2018_MBL_Avalanche}%
  \BibitemOpen
  \bibfield  {author} {\bibinfo {author} {\bibfnamefont {Thimoth\'ee}\
  \bibnamefont {Thiery}}, \bibinfo {author} {\bibfnamefont
  {Fran\ifmmode\mbox{\c{c}}\else\c{c}\fi{}ois}\ \bibnamefont {Huveneers}},
  \bibinfo {author} {\bibfnamefont {Markus}\ \bibnamefont {M\"uller}}, \ and\
  \bibinfo {author} {\bibfnamefont {Wojciech}\ \bibnamefont {De~Roeck}},\
  }\bibfield  {title} {\enquote {\bibinfo {title} {Many-body delocalization as
  a quantum avalanche},}\ }\href {\doibase 10.1103/PhysRevLett.121.140601}
  {\bibfield  {journal} {\bibinfo  {journal} {Phys. Rev. Lett.}\ }\textbf
  {\bibinfo {volume} {121}},\ \bibinfo {pages} {140601} (\bibinfo {year}
  {2018})}\BibitemShut {NoStop}%
\bibitem [{\citenamefont {Goremykina}\ \emph {et~al.}(2019)\citenamefont
  {Goremykina}, \citenamefont {Vasseur},\ and\ \citenamefont
  {Serbyn}}]{Goremykina2019-1}%
  \BibitemOpen
  \bibfield  {author} {\bibinfo {author} {\bibfnamefont {Anna}\ \bibnamefont
  {Goremykina}}, \bibinfo {author} {\bibfnamefont {Romain}\ \bibnamefont
  {Vasseur}}, \ and\ \bibinfo {author} {\bibfnamefont {Maksym}\ \bibnamefont
  {Serbyn}},\ }\bibfield  {title} {\enquote {\bibinfo {title} {Analytically
  solvable renormalization group for the many-body localization transition},}\
  }\href {\doibase 10.1103/PhysRevLett.122.040601} {\bibfield  {journal}
  {\bibinfo  {journal} {Phys. Rev. Lett.}\ }\textbf {\bibinfo {volume} {122}},\
  \bibinfo {pages} {040601} (\bibinfo {year} {2019})}\BibitemShut {NoStop}%
\bibitem [{\citenamefont {Dumitrescu}\ \emph {et~al.}(2019)\citenamefont
  {Dumitrescu}, \citenamefont {Goremykina}, \citenamefont {Parameswaran},
  \citenamefont {Serbyn},\ and\ \citenamefont {Vasseur}}]{Goremykina2019-2}%
  \BibitemOpen
  \bibfield  {author} {\bibinfo {author} {\bibfnamefont {Philipp~T.}\
  \bibnamefont {Dumitrescu}}, \bibinfo {author} {\bibfnamefont {Anna}\
  \bibnamefont {Goremykina}}, \bibinfo {author} {\bibfnamefont {Siddharth~A.}\
  \bibnamefont {Parameswaran}}, \bibinfo {author} {\bibfnamefont {Maksym}\
  \bibnamefont {Serbyn}}, \ and\ \bibinfo {author} {\bibfnamefont {Romain}\
  \bibnamefont {Vasseur}},\ }\bibfield  {title} {\enquote {\bibinfo {title}
  {{K}osterlitz-{T}houless scaling at many-body localization phase
  transitions},}\ }\href {\doibase 10.1103/PhysRevB.99.094205} {\bibfield
  {journal} {\bibinfo  {journal} {Phys. Rev. B}\ }\textbf {\bibinfo {volume}
  {99}},\ \bibinfo {pages} {094205} (\bibinfo {year} {2019})}\BibitemShut
  {NoStop}%
\bibitem [{\citenamefont {Morningstar}\ and\ \citenamefont
  {Huse}(2019{\natexlab{a}})}]{Alan_2019_1}%
  \BibitemOpen
  \bibfield  {author} {\bibinfo {author} {\bibfnamefont {Alan}\ \bibnamefont
  {Morningstar}}\ and\ \bibinfo {author} {\bibfnamefont {David~A.}\
  \bibnamefont {Huse}},\ }\bibfield  {title} {\enquote {\bibinfo {title}
  {Renormalization-group study of the many-body localization transition in one
  dimension},}\ }\href {\doibase 10.1103/PhysRevB.99.224205} {\bibfield
  {journal} {\bibinfo  {journal} {Phys. Rev. B}\ }\textbf {\bibinfo {volume}
  {99}},\ \bibinfo {pages} {224205} (\bibinfo {year}
  {2019}{\natexlab{a}})}\BibitemShut {NoStop}%
\bibitem [{\citenamefont {Morningstar}\ \emph {et~al.}(2020)\citenamefont
  {Morningstar}, \citenamefont {Huse},\ and\ \citenamefont
  {Imbrie}}]{Alan_2020_2}%
  \BibitemOpen
  \bibfield  {author} {\bibinfo {author} {\bibfnamefont {Alan}\ \bibnamefont
  {Morningstar}}, \bibinfo {author} {\bibfnamefont {David~A.}\ \bibnamefont
  {Huse}}, \ and\ \bibinfo {author} {\bibfnamefont {John~Z.}\ \bibnamefont
  {Imbrie}},\ }\bibfield  {title} {\enquote {\bibinfo {title} {Many-body
  localization near the critical point},}\ }\href {\doibase
  10.1103/PhysRevB.102.125134} {\bibfield  {journal} {\bibinfo  {journal}
  {Phys. Rev. B}\ }\textbf {\bibinfo {volume} {102}},\ \bibinfo {pages}
  {125134} (\bibinfo {year} {2020})}\BibitemShut {NoStop}%
\bibitem [{\citenamefont {Morningstar}\ and\ \citenamefont
  {Huse}(2019{\natexlab{b}})}]{Alan_2019_3}%
  \BibitemOpen
  \bibfield  {author} {\bibinfo {author} {\bibfnamefont {Alan}\ \bibnamefont
  {Morningstar}}\ and\ \bibinfo {author} {\bibfnamefont {David~A.}\
  \bibnamefont {Huse}},\ }\bibfield  {title} {\enquote {\bibinfo {title}
  {Renormalization-group study of the many-body localization transition in one
  dimension},}\ }\href {\doibase 10.1103/PhysRevB.99.224205} {\bibfield
  {journal} {\bibinfo  {journal} {Phys. Rev. B}\ }\textbf {\bibinfo {volume}
  {99}},\ \bibinfo {pages} {224205} (\bibinfo {year}
  {2019}{\natexlab{b}})}\BibitemShut {NoStop}%
\bibitem [{\citenamefont {Altshuler}\ \emph {et~al.}(1997)\citenamefont
  {Altshuler}, \citenamefont {Gefen}, \citenamefont {Kamenev},\ and\
  \citenamefont {Levitov}}]{Alt97}%
  \BibitemOpen
  \bibfield  {author} {\bibinfo {author} {\bibfnamefont {Boris~L.}\
  \bibnamefont {Altshuler}}, \bibinfo {author} {\bibfnamefont {Yuval}\
  \bibnamefont {Gefen}}, \bibinfo {author} {\bibfnamefont {Alex}\ \bibnamefont
  {Kamenev}}, \ and\ \bibinfo {author} {\bibfnamefont {Leonid~S.}\ \bibnamefont
  {Levitov}},\ }\bibfield  {title} {\enquote {\bibinfo {title} {Quasiparticle
  lifetime in a finite system: A nonperturbative approach},}\ }\href {\doibase
  10.1103/PhysRevLett.78.2803} {\bibfield  {journal} {\bibinfo  {journal}
  {Phys. Rev. Lett.}\ }\textbf {\bibinfo {volume} {78}},\ \bibinfo {pages}
  {2803--2806} (\bibinfo {year} {1997})}\BibitemShut {NoStop}%
\bibitem [{Note1()}]{Note1}%
  \BibitemOpen
  \bibinfo {note} {The main directions here can be characterized as the mapping
  of the MBL to the localization on hierarchical structures like a random
  regular graph~(see, e.g., \cite
  {Biroli:2012vk,Deluca14,Alt16,TikMir16,Tik16,Biroli2017Dynamics,
  Sonner17,Lemarie17Small_K,
  Biroli2018delocalization,Kra18,parisi2019anderson,bera19,DeTomasi2019Subdiffusion,Tikh2019_K(w),Tikh2019Critical,Biroli_Tarzia2020subdiffusion,tikhonov2020eigenstate}),
  considerations of the spin models with uncorrelated on-site disorder like a
  Quantum Random energy model (see, e.g.,~\cite {
  Laumann2014QREM,Baldwin2016qRem,faoro2019QREM,Smelyanskiy2020QREM,Kechedzhi2018QREM,
  biroli2020QREM} and some other ones~~\cite {Logan19, Basko06, Roy1, Roy2,
  GDT_2020_Fock}), as well as the associating of the MBL transition with the
  ergodic transition in random-matrix models~\cite
  {Kravtsov_NJP2015,Tarzia_2020,LNRP2020_RRG,LNRP2020_WE}.}\BibitemShut {Stop}%
\bibitem [{\citenamefont {Evers}\ and\ \citenamefont
  {Mirlin}(2008)}]{Evers2008Review}%
  \BibitemOpen
  \bibfield  {author} {\bibinfo {author} {\bibfnamefont {Ferdinand}\
  \bibnamefont {Evers}}\ and\ \bibinfo {author} {\bibfnamefont {Alexander~D.}\
  \bibnamefont {Mirlin}},\ }\bibfield  {title} {\enquote {\bibinfo {title}
  {{A}nderson transitions},}\ }\href {\doibase 10.1103/RevModPhys.80.1355}
  {\bibfield  {journal} {\bibinfo  {journal} {Rev. Mod. Phys.}\ }\textbf
  {\bibinfo {volume} {80}},\ \bibinfo {pages} {1355--1417} (\bibinfo {year}
  {2008})}\BibitemShut {NoStop}%
\bibitem [{\citenamefont {Page}(1993)}]{Page1993}%
  \BibitemOpen
  \bibfield  {author} {\bibinfo {author} {\bibfnamefont {Don~N.}\ \bibnamefont
  {Page}},\ }\bibfield  {title} {\enquote {\bibinfo {title} {{Average entropy
  of a subsystem}},}\ }\href {\doibase 10.1103/PhysRevLett.71.1291} {\bibfield
  {journal} {\bibinfo  {journal} {Phys. Rev. Lett.}\ }\textbf {\bibinfo
  {volume} {71}},\ \bibinfo {pages} {1291--1294} (\bibinfo {year}
  {1993})}\BibitemShut {NoStop}%
\bibitem [{WE_()}]{WE_ETH-footnote}%
  \BibitemOpen
  \href@noop {} {}\bibinfo {note} {However, there are several recent
  observations showing some significant deviations from complete randomness at
  the infinite temperature in a generic local non-integrable many-body system
  (see,
  e.g.,~\cite{Haque2015PR_vs_Sent,LuitzBarlev_PRL16,Beugeling_coefficients_PRE2018,HamazakiUeda_atypicality_PRL2018,Baecker2019,Sent2020_Haque}
  and references therein).}\BibitemShut {Stop}%
\bibitem [{\citenamefont {Tikhonov}\ and\ \citenamefont
  {Mirlin}(2018)}]{Tikhonov2018MBL_long-range}%
  \BibitemOpen
  \bibfield  {author} {\bibinfo {author} {\bibfnamefont {K.~S.}\ \bibnamefont
  {Tikhonov}}\ and\ \bibinfo {author} {\bibfnamefont {A.~D.}\ \bibnamefont
  {Mirlin}},\ }\bibfield  {title} {\enquote {\bibinfo {title} {Many-body
  localization transition with power-law interactions: Statistics of
  eigenstates},}\ }\href {\doibase 10.1103/PhysRevB.97.214205} {\bibfield
  {journal} {\bibinfo  {journal} {Phys. Rev. B}\ }\textbf {\bibinfo {volume}
  {97}},\ \bibinfo {pages} {214205} (\bibinfo {year} {2018})}\BibitemShut
  {NoStop}%
\bibitem [{\citenamefont {Mac\'e}\ \emph {et~al.}(2019)\citenamefont {Mac\'e},
  \citenamefont {Alet},\ and\ \citenamefont
  {Laflorencie}}]{Mace_Laflorencie2019_XXZ}%
  \BibitemOpen
  \bibfield  {author} {\bibinfo {author} {\bibfnamefont {Nicolas}\ \bibnamefont
  {Mac\'e}}, \bibinfo {author} {\bibfnamefont {Fabien}\ \bibnamefont {Alet}}, \
  and\ \bibinfo {author} {\bibfnamefont {Nicolas}\ \bibnamefont
  {Laflorencie}},\ }\bibfield  {title} {\enquote {\bibinfo {title}
  {Multifractal scalings across the many-body localization transition},}\
  }\href {\doibase 10.1103/PhysRevLett.123.180601} {\bibfield  {journal}
  {\bibinfo  {journal} {Phys. Rev. Lett.}\ }\textbf {\bibinfo {volume} {123}},\
  \bibinfo {pages} {180601} (\bibinfo {year} {2019})}\BibitemShut {NoStop}%
\bibitem [{\citenamefont {Luitz}\ \emph {et~al.}(2020)\citenamefont {Luitz},
  \citenamefont {Khaymovich},\ and\ \citenamefont
  {Lev}}]{Luitz_Khaymovich_BarLev_multifrac_SciPost2020}%
  \BibitemOpen
  \bibfield  {author} {\bibinfo {author} {\bibfnamefont {David~J.}\
  \bibnamefont {Luitz}}, \bibinfo {author} {\bibfnamefont {Ivan~M.}\
  \bibnamefont {Khaymovich}}, \ and\ \bibinfo {author} {\bibfnamefont
  {Yevgeny~Bar}\ \bibnamefont {Lev}},\ }\bibfield  {title} {\enquote {\bibinfo
  {title} {{Multifractality and its role in anomalous transport in the
  disordered XXZ spin-chain}},}\ }\href {\doibase
  10.21468/SciPostPhysCore.2.2.006} {\bibfield  {journal} {\bibinfo  {journal}
  {SciPost Phys. Core}\ }\textbf {\bibinfo {volume} {2}},\ \bibinfo {pages} {6}
  (\bibinfo {year} {2020})}\BibitemShut {NoStop}%
\bibitem [{\citenamefont {Tarzia}(2020)}]{Tarzia_2020}%
  \BibitemOpen
  \bibfield  {author} {\bibinfo {author} {\bibfnamefont {M.}~\bibnamefont
  {Tarzia}},\ }\bibfield  {title} {\enquote {\bibinfo {title} {Many-body
  localization transition in hilbert space},}\ }\href {\doibase
  10.1103/PhysRevB.102.014208} {\bibfield  {journal} {\bibinfo  {journal}
  {Phys. Rev. B}\ }\textbf {\bibinfo {volume} {102}},\ \bibinfo {pages}
  {014208} (\bibinfo {year} {2020})}\BibitemShut {NoStop}%
\bibitem [{\citenamefont {De~Tomasi}\ and\ \citenamefont
  {Khaymovich}(2020)}]{DeTomasi_2020}%
  \BibitemOpen
  \bibfield  {author} {\bibinfo {author} {\bibfnamefont {Giuseppe}\
  \bibnamefont {De~Tomasi}}\ and\ \bibinfo {author} {\bibfnamefont {Ivan~M.}\
  \bibnamefont {Khaymovich}},\ }\bibfield  {title} {\enquote {\bibinfo {title}
  {Multifractality meets entanglement: Relation for nonergodic extended
  states},}\ }\href {\doibase 10.1103/PhysRevLett.124.200602} {\bibfield
  {journal} {\bibinfo  {journal} {Phys. Rev. Lett.}\ }\textbf {\bibinfo
  {volume} {124}},\ \bibinfo {pages} {200602} (\bibinfo {year}
  {2020})}\BibitemShut {NoStop}%
\bibitem [{\citenamefont {Imbrie}(2016)}]{Imbrie2016}%
  \BibitemOpen
  \bibfield  {author} {\bibinfo {author} {\bibfnamefont {John~Z.}\ \bibnamefont
  {Imbrie}},\ }\bibfield  {title} {\enquote {\bibinfo {title} {{On Many-Body
  Localization for Quantum Spin Chains}},}\ }\href {\doibase
  10.1007/s10955-016-1508-x} {\bibfield  {journal} {\bibinfo  {journal} {J.
  Stat. Phys.}\ }\textbf {\bibinfo {volume} {163}},\ \bibinfo {pages}
  {998--1048} (\bibinfo {year} {2016})}\BibitemShut {NoStop}%
\bibitem [{\citenamefont {{Abanin}}\ \emph {et~al.}(2019)\citenamefont
  {{Abanin}}, \citenamefont {{Bardarson}}, \citenamefont {{De Tomasi}},
  \citenamefont {{Gopalakrishnan}}, \citenamefont {{Khemani}}, \citenamefont
  {{Parameswaran}}, \citenamefont {{Pollmann}}, \citenamefont {{Potter}},
  \citenamefont {{Serbyn}},\ and\ \citenamefont {{Vasseur}}}]{2019Abanin_many}%
  \BibitemOpen
  \bibfield  {author} {\bibinfo {author} {\bibfnamefont {D.~A.}\ \bibnamefont
  {{Abanin}}}, \bibinfo {author} {\bibfnamefont {J.~H.}\ \bibnamefont
  {{Bardarson}}}, \bibinfo {author} {\bibfnamefont {G.}~\bibnamefont {{De
  Tomasi}}}, \bibinfo {author} {\bibfnamefont {S.}~\bibnamefont
  {{Gopalakrishnan}}}, \bibinfo {author} {\bibfnamefont {V.}~\bibnamefont
  {{Khemani}}}, \bibinfo {author} {\bibfnamefont {S.~A.}\ \bibnamefont
  {{Parameswaran}}}, \bibinfo {author} {\bibfnamefont {F.}~\bibnamefont
  {{Pollmann}}}, \bibinfo {author} {\bibfnamefont {A.~C.}\ \bibnamefont
  {{Potter}}}, \bibinfo {author} {\bibfnamefont {M.}~\bibnamefont {{Serbyn}}},
  \ and\ \bibinfo {author} {\bibfnamefont {R.}~\bibnamefont {{Vasseur}}},\
  }\bibfield  {title} {\enquote {\bibinfo {title} {{Distinguishing localization
  from chaos: challenges in finite-size systems}},}\ }\href@noop {} {\bibfield
  {journal} {\bibinfo  {journal} {arXiv e-prints}\ ,\ \bibinfo {eid}
  {arXiv:1911.04501}} (\bibinfo {year} {2019})},\ \Eprint
  {http://arxiv.org/abs/1911.04501} {arXiv:1911.04501 [cond-mat.str-el]}
  \BibitemShut {NoStop}%
\bibitem [{\citenamefont {Derrida}(1981)}]{Derrida1981REM}%
  \BibitemOpen
  \bibfield  {author} {\bibinfo {author} {\bibfnamefont {Bernard}\ \bibnamefont
  {Derrida}},\ }\bibfield  {title} {\enquote {\bibinfo {title} {Random-energy
  model: An exactly solvable model of disordered systems},}\ }\href {\doibase
  10.1103/PhysRevB.24.2613} {\bibfield  {journal} {\bibinfo  {journal} {Phys.
  Rev. B}\ }\textbf {\bibinfo {volume} {24}},\ \bibinfo {pages} {2613--2626}
  (\bibinfo {year} {1981})}\BibitemShut {NoStop}%
\bibitem [{\citenamefont {Goldschmidt}(1990)}]{Gold91}%
  \BibitemOpen
  \bibfield  {author} {\bibinfo {author} {\bibfnamefont {Yadin~Y.}\
  \bibnamefont {Goldschmidt}},\ }\bibfield  {title} {\enquote {\bibinfo {title}
  {Solvable model of the quantum spin glass in a transverse field},}\ }\href
  {\doibase 10.1103/PhysRevB.41.4858} {\bibfield  {journal} {\bibinfo
  {journal} {Phys. Rev. B}\ }\textbf {\bibinfo {volume} {41}},\ \bibinfo
  {pages} {4858--4861} (\bibinfo {year} {1990})}\BibitemShut {NoStop}%
\bibitem [{\citenamefont {{Manai}}\ and\ \citenamefont
  {{Warzel}}(2020)}]{Manai20}%
  \BibitemOpen
  \bibfield  {author} {\bibinfo {author} {\bibfnamefont {Chokri}\ \bibnamefont
  {{Manai}}}\ and\ \bibinfo {author} {\bibfnamefont {Simone}\ \bibnamefont
  {{Warzel}}},\ }\bibfield  {title} {\enquote {\bibinfo {title} {{Generalized
  Random Energy Models in a Transversal Magnetic Field: Free Energy and Phase
  Diagrams}},}\ }\href@noop {} {\bibfield  {journal} {\bibinfo  {journal}
  {arXiv e-prints}\ ,\ \bibinfo {eid} {arXiv:2007.03290}} (\bibinfo {year}
  {2020})},\ \Eprint {http://arxiv.org/abs/2007.03290} {arXiv:2007.03290
  [math-ph]} \BibitemShut {NoStop}%
\bibitem [{\citenamefont {Aizenman}\ and\ \citenamefont
  {Warzel}(2015)}]{AizWar15}%
  \BibitemOpen
  \bibfield  {author} {\bibinfo {author} {\bibfnamefont {M.}~\bibnamefont
  {Aizenman}}\ and\ \bibinfo {author} {\bibfnamefont {S.}~\bibnamefont
  {Warzel}},\ }\bibfield  {title} {\enquote {\bibinfo {title} {Random
  operators: Disorder effects on quantum spectra and dynamics},}\ }\href@noop
  {} {\  (\bibinfo {year} {2015})}\BibitemShut {NoStop}%
\bibitem [{\citenamefont {Torres-Herrera}\ \emph {et~al.}(2020)\citenamefont
  {Torres-Herrera}, \citenamefont {De~Tomasi}, \citenamefont {Schiulaz},
  \citenamefont {P\'erez-Bernal},\ and\ \citenamefont {Santos}}]{GDT_2020_Lea}%
  \BibitemOpen
  \bibfield  {author} {\bibinfo {author} {\bibfnamefont {E.~Jonathan}\
  \bibnamefont {Torres-Herrera}}, \bibinfo {author} {\bibfnamefont {Giuseppe}\
  \bibnamefont {De~Tomasi}}, \bibinfo {author} {\bibfnamefont {Mauro}\
  \bibnamefont {Schiulaz}}, \bibinfo {author} {\bibfnamefont {Francisco}\
  \bibnamefont {P\'erez-Bernal}}, \ and\ \bibinfo {author} {\bibfnamefont
  {Lea~F.}\ \bibnamefont {Santos}},\ }\bibfield  {title} {\enquote {\bibinfo
  {title} {Self-averaging in many-body quantum systems out of equilibrium:
  Approach to the localized phase},}\ }\href {\doibase
  10.1103/PhysRevB.102.094310} {\bibfield  {journal} {\bibinfo  {journal}
  {Phys. Rev. B}\ }\textbf {\bibinfo {volume} {102}},\ \bibinfo {pages}
  {094310} (\bibinfo {year} {2020})}\BibitemShut {NoStop}%
\bibitem [{Note2()}]{Note2}%
  \BibitemOpen
  \bibinfo {note} {The following properties hold $\Pi (x)\ge 0$ and $\DOTSB
  \sum@ \slimits@ _x \Pi (x)=1$.}\BibitemShut {Stop}%
\bibitem [{dS_()}]{dS_footnote}%
  \BibitemOpen
  \href@noop {} {}\bibinfo {note} {The factor $2$ in the discrete derivative
  $\partial S$ of the entanglement entropy normalizes it to the ergodic
  value.}\BibitemShut {Stop}%
\bibitem [{\citenamefont {Laumann}\ \emph {et~al.}(2014)\citenamefont
  {Laumann}, \citenamefont {Pal},\ and\ \citenamefont
  {Scardicchio}}]{Laumann2014QREM}%
  \BibitemOpen
  \bibfield  {author} {\bibinfo {author} {\bibfnamefont {C.~R.}\ \bibnamefont
  {Laumann}}, \bibinfo {author} {\bibfnamefont {A.}~\bibnamefont {Pal}}, \ and\
  \bibinfo {author} {\bibfnamefont {A.}~\bibnamefont {Scardicchio}},\
  }\bibfield  {title} {\enquote {\bibinfo {title} {Many-body mobility edge in a
  mean-field quantum spin glass},}\ }\href {\doibase
  10.1103/PhysRevLett.113.200405} {\bibfield  {journal} {\bibinfo  {journal}
  {Phys. Rev. Lett.}\ }\textbf {\bibinfo {volume} {113}},\ \bibinfo {pages}
  {200405} (\bibinfo {year} {2014})}\BibitemShut {NoStop}%
\bibitem [{\citenamefont {Baldwin}\ \emph {et~al.}(2016)\citenamefont
  {Baldwin}, \citenamefont {Laumann}, \citenamefont {Pal},\ and\ \citenamefont
  {Scardicchio}}]{Baldwin2016qRem}%
  \BibitemOpen
  \bibfield  {author} {\bibinfo {author} {\bibfnamefont {C.~L.}\ \bibnamefont
  {Baldwin}}, \bibinfo {author} {\bibfnamefont {C.~R.}\ \bibnamefont
  {Laumann}}, \bibinfo {author} {\bibfnamefont {A.}~\bibnamefont {Pal}}, \ and\
  \bibinfo {author} {\bibfnamefont {A.}~\bibnamefont {Scardicchio}},\
  }\bibfield  {title} {\enquote {\bibinfo {title} {{The many-body localized
  phase of the quantum random energy model}},}\ }\href@noop {} {\bibfield
  {journal} {\bibinfo  {journal} {Phys. Rev. B}\ }\textbf {\bibinfo {volume}
  {93}},\ \bibinfo {pages} {024202} (\bibinfo {year} {2016})}\BibitemShut
  {NoStop}%
\bibitem [{\citenamefont {Faoro}\ \emph {et~al.}(2019)\citenamefont {Faoro},
  \citenamefont {Feigel’man},\ and\ \citenamefont {Ioffe}}]{faoro2019QREM}%
  \BibitemOpen
  \bibfield  {author} {\bibinfo {author} {\bibfnamefont {Lara}\ \bibnamefont
  {Faoro}}, \bibinfo {author} {\bibfnamefont {Mikhail~V}\ \bibnamefont
  {Feigel’man}}, \ and\ \bibinfo {author} {\bibfnamefont {Lev}\ \bibnamefont
  {Ioffe}},\ }\bibfield  {title} {\enquote {\bibinfo {title} {Non-ergodic
  extended phase of the quantum random energy model},}\ }\href@noop {}
  {\bibfield  {journal} {\bibinfo  {journal} {Annals of Physics}\ }\textbf
  {\bibinfo {volume} {409}},\ \bibinfo {pages} {167916} (\bibinfo {year}
  {2019})}\BibitemShut {NoStop}%
\bibitem [{\citenamefont {Smelyanskiy}\ \emph {et~al.}(2020)\citenamefont
  {Smelyanskiy}, \citenamefont {Kechedzhi}, \citenamefont {Boixo},
  \citenamefont {Isakov}, \citenamefont {Neven},\ and\ \citenamefont
  {Altshuler}}]{Smelyanskiy2020QREM}%
  \BibitemOpen
  \bibfield  {author} {\bibinfo {author} {\bibfnamefont {Vadim~N.}\
  \bibnamefont {Smelyanskiy}}, \bibinfo {author} {\bibfnamefont {Kostyantyn}\
  \bibnamefont {Kechedzhi}}, \bibinfo {author} {\bibfnamefont {Sergio}\
  \bibnamefont {Boixo}}, \bibinfo {author} {\bibfnamefont {Sergei~V.}\
  \bibnamefont {Isakov}}, \bibinfo {author} {\bibfnamefont {Hartmut}\
  \bibnamefont {Neven}}, \ and\ \bibinfo {author} {\bibfnamefont {Boris}\
  \bibnamefont {Altshuler}},\ }\bibfield  {title} {\enquote {\bibinfo {title}
  {Nonergodic delocalized states for efficient population transfer within a
  narrow band of the energy landscape},}\ }\href {\doibase
  10.1103/PhysRevX.10.011017} {\bibfield  {journal} {\bibinfo  {journal} {Phys.
  Rev. X}\ }\textbf {\bibinfo {volume} {10}},\ \bibinfo {pages} {011017}
  (\bibinfo {year} {2020})}\BibitemShut {NoStop}%
\bibitem [{\citenamefont {Kechedzhi}\ \emph {et~al.}()\citenamefont
  {Kechedzhi}, \citenamefont {Smelyanskiy}, \citenamefont {McClean},
  \citenamefont {Denchev}, \citenamefont {Mohseni}, \citenamefont {Isakov},
  \citenamefont {Boixo}, \citenamefont {Altshuler},\ and\ \citenamefont
  {Neven}}]{Kechedzhi2018QREM}%
  \BibitemOpen
  \bibfield  {author} {\bibinfo {author} {\bibfnamefont {K.}~\bibnamefont
  {Kechedzhi}}, \bibinfo {author} {\bibfnamefont {V.~N.}\ \bibnamefont
  {Smelyanskiy}}, \bibinfo {author} {\bibfnamefont {J.~R}\ \bibnamefont
  {McClean}}, \bibinfo {author} {\bibfnamefont {V.~S}\ \bibnamefont {Denchev}},
  \bibinfo {author} {\bibfnamefont {M.}~\bibnamefont {Mohseni}}, \bibinfo
  {author} {\bibfnamefont {S.~V.}\ \bibnamefont {Isakov}}, \bibinfo {author}
  {\bibfnamefont {S.}~\bibnamefont {Boixo}}, \bibinfo {author} {\bibfnamefont
  {B.~L.}\ \bibnamefont {Altshuler}}, \ and\ \bibinfo {author} {\bibfnamefont
  {H.}~\bibnamefont {Neven}},\ }\href@noop {} {\enquote {\bibinfo {title}
  {Efficient population transfer via non-ergodic extended states in quantum
  spin glass},}\ }\BibitemShut {NoStop}%
\bibitem [{\citenamefont {Aubry}\ and\ \citenamefont {André}(1980)}]{Aubry80}%
  \BibitemOpen
  \bibfield  {author} {\bibinfo {author} {\bibfnamefont {Serge}\ \bibnamefont
  {Aubry}}\ and\ \bibinfo {author} {\bibfnamefont {Gilles}\ \bibnamefont
  {André}},\ }\bibfield  {title} {\enquote {\bibinfo {title} {Analyticity
  breaking and {A}nderson localization in incommensurate lattices},}\
  }\href@noop {} {\bibfield  {journal} {\bibinfo  {journal} {Proceedings, VIII
  International Colloquium on Group-Theoretical Methods in Physics}\ }\textbf
  {\bibinfo {volume} {3}} (\bibinfo {year} {1980})}\BibitemShut {NoStop}%
\bibitem [{\citenamefont {{{\v{S}}untajs}}\ \emph {et~al.}(2019)\citenamefont
  {{{\v{S}}untajs}}, \citenamefont {{Bon{\v{c}}a}}, \citenamefont {{Prosen}},\
  and\ \citenamefont {{Vidmar}}}]{Lev19}%
  \BibitemOpen
  \bibfield  {author} {\bibinfo {author} {\bibfnamefont {J.}~\bibnamefont
  {{{\v{S}}untajs}}}, \bibinfo {author} {\bibfnamefont {J.}~\bibnamefont
  {{Bon{\v{c}}a}}}, \bibinfo {author} {\bibfnamefont {T.}~\bibnamefont
  {{Prosen}}}, \ and\ \bibinfo {author} {\bibfnamefont {L.}~\bibnamefont
  {{Vidmar}}},\ }\bibfield  {title} {\enquote {\bibinfo {title} {{Quantum chaos
  challenges many-body localization}},}\ }\href@noop {} {\bibfield  {journal}
  {\bibinfo  {journal} {arXiv e-prints}\ ,\ \bibinfo {eid} {arXiv:1905.06345}}
  (\bibinfo {year} {2019})},\ \Eprint {http://arxiv.org/abs/1905.06345}
  {arXiv:1905.06345 [cond-mat.str-el]} \BibitemShut {NoStop}%
\bibitem [{\citenamefont {{Sels}}\ and\ \citenamefont
  {{Polkovnikov}}(2020)}]{Sels_2020}%
  \BibitemOpen
  \bibfield  {author} {\bibinfo {author} {\bibfnamefont {Dries}\ \bibnamefont
  {{Sels}}}\ and\ \bibinfo {author} {\bibfnamefont {Anatoli}\ \bibnamefont
  {{Polkovnikov}}},\ }\bibfield  {title} {\enquote {\bibinfo {title}
  {{Dynamical obstruction to localization in a disordered spin chain}},}\
  }\href@noop {} {\bibfield  {journal} {\bibinfo  {journal} {arXiv e-prints}\
  ,\ \bibinfo {eid} {arXiv:2009.04501}} (\bibinfo {year} {2020})},\ \Eprint
  {http://arxiv.org/abs/2009.04501} {arXiv:2009.04501 [quant-ph]} \BibitemShut
  {NoStop}%
\bibitem [{Note3()}]{Note3}%
  \BibitemOpen
  \bibinfo {note} {Repeated indices are summed.}\BibitemShut {Stop}%
\bibitem [{Note4()}]{Note4}%
  \BibitemOpen
  \bibinfo {note} {Cf. the Supplementary material of~\cite
  {Mace_Laflorencie2019_XXZ}}\BibitemShut {NoStop}%
\bibitem [{\citenamefont {Avila}\ and\ \citenamefont
  {Jitomirskaya}(2009)}]{AviJit2009}%
  \BibitemOpen
  \bibfield  {author} {\bibinfo {author} {\bibfnamefont {A.}~\bibnamefont
  {Avila}}\ and\ \bibinfo {author} {\bibfnamefont {S.}~\bibnamefont
  {Jitomirskaya}},\ }\bibfield  {title} {\enquote {\bibinfo {title} {{The Ten
  {M}artini Problem}},}\ }\href {\doibase 10.4007/annals.2009.170.303}
  {\bibfield  {journal} {\bibinfo  {journal} {Ann. Math.}\ }\textbf {\bibinfo
  {volume} {170}},\ \bibinfo {pages} {303--342} (\bibinfo {year}
  {2009})}\BibitemShut {NoStop}%
\bibitem [{\citenamefont {Biroli}\ \emph {et~al.}(2012)\citenamefont {Biroli},
  \citenamefont {Ribeiro-Teixeira},\ and\ \citenamefont
  {Tarzia}}]{Biroli:2012vk}%
  \BibitemOpen
  \bibfield  {author} {\bibinfo {author} {\bibfnamefont {G}~\bibnamefont
  {Biroli}}, \bibinfo {author} {\bibfnamefont {A~C}\ \bibnamefont
  {Ribeiro-Teixeira}}, \ and\ \bibinfo {author} {\bibfnamefont {M}~\bibnamefont
  {Tarzia}},\ }\bibfield  {title} {\enquote {\bibinfo {title} {Difference
  between level statistics, ergodicity and localization transitions on the
  {Bethe} lattice},}\ }\href {https://arxiv.org/abs/1211.7334} {\bibfield
  {journal} {\bibinfo  {journal} {arXiv:1211.7334 [cond-mat.dis-nn]}\ }
  (\bibinfo {year} {2012})}\BibitemShut {NoStop}%
\bibitem [{\citenamefont {De~Luca}\ \emph {et~al.}(2014)\citenamefont
  {De~Luca}, \citenamefont {Altshuler}, \citenamefont {Kravtsov},\ and\
  \citenamefont {Scardicchio}}]{Deluca14}%
  \BibitemOpen
  \bibfield  {author} {\bibinfo {author} {\bibfnamefont {A.}~\bibnamefont
  {De~Luca}}, \bibinfo {author} {\bibfnamefont {B.~L.}\ \bibnamefont
  {Altshuler}}, \bibinfo {author} {\bibfnamefont {V.~E.}\ \bibnamefont
  {Kravtsov}}, \ and\ \bibinfo {author} {\bibfnamefont {A.}~\bibnamefont
  {Scardicchio}},\ }\bibfield  {title} {\enquote {\bibinfo {title} {{A}nderson
  localization on the bethe lattice: Nonergodicity of extended states},}\
  }\href {\doibase 10.1103/PhysRevLett.113.046806} {\bibfield  {journal}
  {\bibinfo  {journal} {Phys. Rev. Lett.}\ }\textbf {\bibinfo {volume} {113}},\
  \bibinfo {pages} {046806} (\bibinfo {year} {2014})}\BibitemShut {NoStop}%
\bibitem [{\citenamefont {Altshuler}\ \emph {et~al.}(2016)\citenamefont
  {Altshuler}, \citenamefont {Cuevas}, \citenamefont {Ioffe},\ and\
  \citenamefont {Kravtsov}}]{Alt16}%
  \BibitemOpen
  \bibfield  {author} {\bibinfo {author} {\bibfnamefont {B.~L.}\ \bibnamefont
  {Altshuler}}, \bibinfo {author} {\bibfnamefont {E.}~\bibnamefont {Cuevas}},
  \bibinfo {author} {\bibfnamefont {L.~B.}\ \bibnamefont {Ioffe}}, \ and\
  \bibinfo {author} {\bibfnamefont {V.~E.}\ \bibnamefont {Kravtsov}},\
  }\bibfield  {title} {\enquote {\bibinfo {title} {Nonergodic phases in
  strongly disordered random regular graphs},}\ }\href {\doibase
  10.1103/PhysRevLett.117.156601} {\bibfield  {journal} {\bibinfo  {journal}
  {Phys. Rev. Lett.}\ }\textbf {\bibinfo {volume} {117}},\ \bibinfo {pages}
  {156601} (\bibinfo {year} {2016})}\BibitemShut {NoStop}%
\bibitem [{\citenamefont {Tikhonov}\ and\ \citenamefont
  {Mirlin}(2016)}]{TikMir16}%
  \BibitemOpen
  \bibfield  {author} {\bibinfo {author} {\bibfnamefont {K.~S.}\ \bibnamefont
  {Tikhonov}}\ and\ \bibinfo {author} {\bibfnamefont {A.~D.}\ \bibnamefont
  {Mirlin}},\ }\bibfield  {title} {\enquote {\bibinfo {title} {Fractality of
  wave functions on a cayley tree: Difference between tree and locally treelike
  graph without boundary},}\ }\href {\doibase 10.1103/PhysRevB.94.184203}
  {\bibfield  {journal} {\bibinfo  {journal} {Phys. Rev. B}\ }\textbf {\bibinfo
  {volume} {94}},\ \bibinfo {pages} {184203} (\bibinfo {year}
  {2016})}\BibitemShut {NoStop}%
\bibitem [{\citenamefont {Tikhonov}\ \emph {et~al.}(2016)\citenamefont
  {Tikhonov}, \citenamefont {Mirlin},\ and\ \citenamefont {Skvortsov}}]{Tik16}%
  \BibitemOpen
  \bibfield  {author} {\bibinfo {author} {\bibfnamefont {K.~S.}\ \bibnamefont
  {Tikhonov}}, \bibinfo {author} {\bibfnamefont {A.~D.}\ \bibnamefont
  {Mirlin}}, \ and\ \bibinfo {author} {\bibfnamefont {M.~A.}\ \bibnamefont
  {Skvortsov}},\ }\bibfield  {title} {\enquote {\bibinfo {title} {{A}nderson
  localization and ergodicity on random regular graphs},}\ }\href {\doibase
  10.1103/PhysRevB.94.220203} {\bibfield  {journal} {\bibinfo  {journal} {Phys.
  Rev. B}\ }\textbf {\bibinfo {volume} {94}},\ \bibinfo {pages} {220203}
  (\bibinfo {year} {2016})}\BibitemShut {NoStop}%
\bibitem [{\citenamefont {Biroli}\ and\ \citenamefont
  {Tarzia}(2017)}]{Biroli2017Dynamics}%
  \BibitemOpen
  \bibfield  {author} {\bibinfo {author} {\bibfnamefont {G.}~\bibnamefont
  {Biroli}}\ and\ \bibinfo {author} {\bibfnamefont {M.}~\bibnamefont
  {Tarzia}},\ }\bibfield  {title} {\enquote {\bibinfo {title} {Delocalized
  glassy dynamics and many-body localization},}\ }\href {\doibase
  10.1103/PhysRevB.96.201114} {\bibfield  {journal} {\bibinfo  {journal} {Phys.
  Rev. B}\ }\textbf {\bibinfo {volume} {96}},\ \bibinfo {pages} {201114}
  (\bibinfo {year} {2017})}\BibitemShut {NoStop}%
\bibitem [{\citenamefont {Sonner}\ \emph {et~al.}(2017)\citenamefont {Sonner},
  \citenamefont {Tikhonov},\ and\ \citenamefont {Mirlin}}]{Sonner17}%
  \BibitemOpen
  \bibfield  {author} {\bibinfo {author} {\bibfnamefont {M.}~\bibnamefont
  {Sonner}}, \bibinfo {author} {\bibfnamefont {K.~S.}\ \bibnamefont
  {Tikhonov}}, \ and\ \bibinfo {author} {\bibfnamefont {A.~D.}\ \bibnamefont
  {Mirlin}},\ }\bibfield  {title} {\enquote {\bibinfo {title} {Multifractality
  of wave functions on a cayley tree: From root to leaves},}\ }\href {\doibase
  10.1103/PhysRevB.96.214204} {\bibfield  {journal} {\bibinfo  {journal} {Phys.
  Rev. B}\ }\textbf {\bibinfo {volume} {96}},\ \bibinfo {pages} {214204}
  (\bibinfo {year} {2017})}\BibitemShut {NoStop}%
\bibitem [{\citenamefont {Garc\'{\i}a-Mata}\ \emph {et~al.}(2017)\citenamefont
  {Garc\'{\i}a-Mata}, \citenamefont {Giraud}, \citenamefont {Georgeot},
  \citenamefont {Martin}, \citenamefont {Dubertrand},\ and\ \citenamefont
  {Lemari\'e}}]{Lemarie17Small_K}%
  \BibitemOpen
  \bibfield  {author} {\bibinfo {author} {\bibfnamefont {I.}~\bibnamefont
  {Garc\'{\i}a-Mata}}, \bibinfo {author} {\bibfnamefont {O.}~\bibnamefont
  {Giraud}}, \bibinfo {author} {\bibfnamefont {B.}~\bibnamefont {Georgeot}},
  \bibinfo {author} {\bibfnamefont {J.}~\bibnamefont {Martin}}, \bibinfo
  {author} {\bibfnamefont {R.}~\bibnamefont {Dubertrand}}, \ and\ \bibinfo
  {author} {\bibfnamefont {G.}~\bibnamefont {Lemari\'e}},\ }\bibfield  {title}
  {\enquote {\bibinfo {title} {Scaling theory of the {A}nderson transition in
  random graphs: Ergodicity and universality},}\ }\href {\doibase
  10.1103/PhysRevLett.118.166801} {\bibfield  {journal} {\bibinfo  {journal}
  {Phys. Rev. Lett.}\ }\textbf {\bibinfo {volume} {118}},\ \bibinfo {pages}
  {166801} (\bibinfo {year} {2017})}\BibitemShut {NoStop}%
\bibitem [{\citenamefont {Biroli}\ and\ \citenamefont
  {Tarzia}(2018)}]{Biroli2018delocalization}%
  \BibitemOpen
  \bibfield  {author} {\bibinfo {author} {\bibfnamefont {Giulio}\ \bibnamefont
  {Biroli}}\ and\ \bibinfo {author} {\bibfnamefont {Marco}\ \bibnamefont
  {Tarzia}},\ }\href@noop {} {\enquote {\bibinfo {title} {Delocalization and
  ergodicity of the {A}nderson model on bethe lattices},}\ } (\bibinfo {year}
  {2018}),\ \Eprint {http://arxiv.org/abs/1810.07545} {arXiv:1810.07545}
  \BibitemShut {NoStop}%
\bibitem [{\citenamefont {Kravtsov}\ \emph {et~al.}(2018)\citenamefont
  {Kravtsov}, \citenamefont {Altshuler},\ and\ \citenamefont {Ioffe}}]{Kra18}%
  \BibitemOpen
  \bibfield  {author} {\bibinfo {author} {\bibfnamefont {V.E.}\ \bibnamefont
  {Kravtsov}}, \bibinfo {author} {\bibfnamefont {B.L.}\ \bibnamefont
  {Altshuler}}, \ and\ \bibinfo {author} {\bibfnamefont {L.B.}\ \bibnamefont
  {Ioffe}},\ }\bibfield  {title} {\enquote {\bibinfo {title} {Non-ergodic
  delocalized phase in {A}nderson model on bethe lattice and regular graph},}\
  }\href {\doibase https://doi.org/10.1016/j.aop.2017.12.009} {\bibfield
  {journal} {\bibinfo  {journal} {Annals of Physics}\ }\textbf {\bibinfo
  {volume} {389}},\ \bibinfo {pages} {148 -- 191} (\bibinfo {year}
  {2018})}\BibitemShut {NoStop}%
\bibitem [{\citenamefont {Parisi}\ \emph {et~al.}(2019)\citenamefont {Parisi},
  \citenamefont {Pascazio}, \citenamefont {Pietracaprina}, \citenamefont
  {Ros},\ and\ \citenamefont {Scardicchio}}]{parisi2019anderson}%
  \BibitemOpen
  \bibfield  {author} {\bibinfo {author} {\bibfnamefont {Giorgio}\ \bibnamefont
  {Parisi}}, \bibinfo {author} {\bibfnamefont {Saverio}\ \bibnamefont
  {Pascazio}}, \bibinfo {author} {\bibfnamefont {Francesca}\ \bibnamefont
  {Pietracaprina}}, \bibinfo {author} {\bibfnamefont {Valentina}\ \bibnamefont
  {Ros}}, \ and\ \bibinfo {author} {\bibfnamefont {Antonello}\ \bibnamefont
  {Scardicchio}},\ }\bibfield  {title} {\enquote {\bibinfo {title} {{A}nderson
  transition on the bethe lattice: an approach with real energies},}\ }\href
  {\doibase 10.1088/1751-8121/ab56e8} {\bibfield  {journal} {\bibinfo
  {journal} {Journal of Physics A: Mathematical and Theoretical}\ }\textbf
  {\bibinfo {volume} {53}},\ \bibinfo {pages} {014003} (\bibinfo {year}
  {2019})}\BibitemShut {NoStop}%
\bibitem [{\citenamefont {Bera}\ \emph {et~al.}(2018)\citenamefont {Bera},
  \citenamefont {De~Tomasi}, \citenamefont {Khaymovich},\ and\ \citenamefont
  {Scardicchio}}]{bera19}%
  \BibitemOpen
  \bibfield  {author} {\bibinfo {author} {\bibfnamefont {Soumya}\ \bibnamefont
  {Bera}}, \bibinfo {author} {\bibfnamefont {Giuseppe}\ \bibnamefont
  {De~Tomasi}}, \bibinfo {author} {\bibfnamefont {Ivan~M.}\ \bibnamefont
  {Khaymovich}}, \ and\ \bibinfo {author} {\bibfnamefont {Antonello}\
  \bibnamefont {Scardicchio}},\ }\bibfield  {title} {\enquote {\bibinfo {title}
  {Return probability for the {A}nderson model on the random regular graph},}\
  }\href {\doibase 10.1103/PhysRevB.98.134205} {\bibfield  {journal} {\bibinfo
  {journal} {Phys. Rev. B}\ }\textbf {\bibinfo {volume} {98}},\ \bibinfo
  {pages} {134205} (\bibinfo {year} {2018})}\BibitemShut {NoStop}%
\bibitem [{\citenamefont {De~Tomasi}\ \emph {et~al.}(2020)\citenamefont
  {De~Tomasi}, \citenamefont {Bera}, \citenamefont {Scardicchio},\ and\
  \citenamefont {Khaymovich}}]{DeTomasi2019Subdiffusion}%
  \BibitemOpen
  \bibfield  {author} {\bibinfo {author} {\bibfnamefont {Giuseppe}\
  \bibnamefont {De~Tomasi}}, \bibinfo {author} {\bibfnamefont {Soumya}\
  \bibnamefont {Bera}}, \bibinfo {author} {\bibfnamefont {Antonello}\
  \bibnamefont {Scardicchio}}, \ and\ \bibinfo {author} {\bibfnamefont
  {Ivan~M.}\ \bibnamefont {Khaymovich}},\ }\bibfield  {title} {\enquote
  {\bibinfo {title} {Subdiffusion in the {A}nderson model on the random regular
  graph},}\ }\href {\doibase 10.1103/PhysRevB.101.100201} {\bibfield  {journal}
  {\bibinfo  {journal} {Phys. Rev. B}\ }\textbf {\bibinfo {volume} {101}},\
  \bibinfo {pages} {100201} (\bibinfo {year} {2020})}\BibitemShut {NoStop}%
\bibitem [{\citenamefont {Tikhonov}\ and\ \citenamefont
  {Mirlin}(2019{\natexlab{a}})}]{Tikh2019_K(w)}%
  \BibitemOpen
  \bibfield  {author} {\bibinfo {author} {\bibfnamefont {K.~S.}\ \bibnamefont
  {Tikhonov}}\ and\ \bibinfo {author} {\bibfnamefont {A.~D.}\ \bibnamefont
  {Mirlin}},\ }\bibfield  {title} {\enquote {\bibinfo {title} {Statistics of
  eigenstates near the localization transition on random regular graphs},}\
  }\href {\doibase 10.1103/PhysRevB.99.024202} {\bibfield  {journal} {\bibinfo
  {journal} {Phys. Rev. B}\ }\textbf {\bibinfo {volume} {99}},\ \bibinfo
  {pages} {024202} (\bibinfo {year} {2019}{\natexlab{a}})}\BibitemShut
  {NoStop}%
\bibitem [{\citenamefont {Tikhonov}\ and\ \citenamefont
  {Mirlin}(2019{\natexlab{b}})}]{Tikh2019Critical}%
  \BibitemOpen
  \bibfield  {author} {\bibinfo {author} {\bibfnamefont {K.~S.}\ \bibnamefont
  {Tikhonov}}\ and\ \bibinfo {author} {\bibfnamefont {A.~D.}\ \bibnamefont
  {Mirlin}},\ }\bibfield  {title} {\enquote {\bibinfo {title} {Critical
  behavior at the localization transition on random regular graphs},}\ }\href
  {\doibase 10.1103/PhysRevB.99.214202} {\bibfield  {journal} {\bibinfo
  {journal} {Phys. Rev. B}\ }\textbf {\bibinfo {volume} {99}},\ \bibinfo
  {pages} {214202} (\bibinfo {year} {2019}{\natexlab{b}})}\BibitemShut
  {NoStop}%
\bibitem [{\citenamefont {Biroli}\ and\ \citenamefont
  {Tarzia}(2020)}]{Biroli_Tarzia2020subdiffusion}%
  \BibitemOpen
  \bibfield  {author} {\bibinfo {author} {\bibfnamefont {G.}~\bibnamefont
  {Biroli}}\ and\ \bibinfo {author} {\bibfnamefont {M.}~\bibnamefont
  {Tarzia}},\ }\bibfield  {title} {\enquote {\bibinfo {title} {Anomalous
  dynamics on the ergodic side of the many-body localization transition and the
  glassy phase of directed polymers in random media},}\ }\href {\doibase
  10.1103/PhysRevB.102.064211} {\bibfield  {journal} {\bibinfo  {journal}
  {Phys. Rev. B}\ }\textbf {\bibinfo {volume} {102}},\ \bibinfo {pages}
  {064211} (\bibinfo {year} {2020})}\BibitemShut {NoStop}%
\bibitem [{\citenamefont {Tikhonov}\ and\ \citenamefont
  {Mirlin}(2020)}]{tikhonov2020eigenstate}%
  \BibitemOpen
  \bibfield  {author} {\bibinfo {author} {\bibfnamefont {Konstantin~S}\
  \bibnamefont {Tikhonov}}\ and\ \bibinfo {author} {\bibfnamefont
  {Alexander~D}\ \bibnamefont {Mirlin}},\ }\href@noop {} {\enquote {\bibinfo
  {title} {Eigenstate correlations around many-body localization transition},}\
  } (\bibinfo {year} {2020}),\ \Eprint {http://arxiv.org/abs/2009.09685}
  {arXiv:2009.09685} \BibitemShut {NoStop}%
\bibitem [{\citenamefont {Biroli}\ \emph {et~al.}(2020)\citenamefont {Biroli},
  \citenamefont {Facoetti}, \citenamefont {Schiro{\'o}}, \citenamefont
  {Tarzia},\ and\ \citenamefont {Vivo}}]{biroli2020QREM}%
  \BibitemOpen
  \bibfield  {author} {\bibinfo {author} {\bibfnamefont {Giulio}\ \bibnamefont
  {Biroli}}, \bibinfo {author} {\bibfnamefont {Davide}\ \bibnamefont
  {Facoetti}}, \bibinfo {author} {\bibfnamefont {Marco}\ \bibnamefont
  {Schiro{\'o}}}, \bibinfo {author} {\bibfnamefont {Marco}\ \bibnamefont
  {Tarzia}}, \ and\ \bibinfo {author} {\bibfnamefont {Pierpaolo}\ \bibnamefont
  {Vivo}},\ }\href@noop {} {\enquote {\bibinfo {title} {Out of equilibrium
  phase diagram of the quantum random energy model},}\ } (\bibinfo {year}
  {2020}),\ \Eprint {http://arxiv.org/abs/2009.09817} {arXiv:2009.09817}
  \BibitemShut {NoStop}%
\bibitem [{\citenamefont {Logan}\ and\ \citenamefont {Welsh}(2019)}]{Logan19}%
  \BibitemOpen
  \bibfield  {author} {\bibinfo {author} {\bibfnamefont {David~E.}\
  \bibnamefont {Logan}}\ and\ \bibinfo {author} {\bibfnamefont {Staszek}\
  \bibnamefont {Welsh}},\ }\bibfield  {title} {\enquote {\bibinfo {title}
  {Many-body localization in {F}ock space: A local perspective},}\ }\href
  {\doibase 10.1103/PhysRevB.99.045131} {\bibfield  {journal} {\bibinfo
  {journal} {Phys. Rev. B}\ }\textbf {\bibinfo {volume} {99}},\ \bibinfo
  {pages} {045131} (\bibinfo {year} {2019})}\BibitemShut {NoStop}%
\bibitem [{\citenamefont {Roy}\ \emph {et~al.}(2019{\natexlab{a}})\citenamefont
  {Roy}, \citenamefont {Logan},\ and\ \citenamefont {Chalker}}]{Roy1}%
  \BibitemOpen
  \bibfield  {author} {\bibinfo {author} {\bibfnamefont {Sthitadhi}\
  \bibnamefont {Roy}}, \bibinfo {author} {\bibfnamefont {David~E.}\
  \bibnamefont {Logan}}, \ and\ \bibinfo {author} {\bibfnamefont {J.~T.}\
  \bibnamefont {Chalker}},\ }\bibfield  {title} {\enquote {\bibinfo {title}
  {Exact solution of a percolation analog for the many-body localization
  transition},}\ }\href {\doibase 10.1103/PhysRevB.99.220201} {\bibfield
  {journal} {\bibinfo  {journal} {Phys. Rev. B}\ }\textbf {\bibinfo {volume}
  {99}},\ \bibinfo {pages} {220201} (\bibinfo {year}
  {2019}{\natexlab{a}})}\BibitemShut {NoStop}%
\bibitem [{\citenamefont {Roy}\ \emph {et~al.}(2019{\natexlab{b}})\citenamefont
  {Roy}, \citenamefont {Chalker},\ and\ \citenamefont {Logan}}]{Roy2}%
  \BibitemOpen
  \bibfield  {author} {\bibinfo {author} {\bibfnamefont {Sthitadhi}\
  \bibnamefont {Roy}}, \bibinfo {author} {\bibfnamefont {J.~T.}\ \bibnamefont
  {Chalker}}, \ and\ \bibinfo {author} {\bibfnamefont {David~E.}\ \bibnamefont
  {Logan}},\ }\bibfield  {title} {\enquote {\bibinfo {title} {Percolation in
  {F}ock space as a proxy for many-body localization},}\ }\href {\doibase
  10.1103/PhysRevB.99.104206} {\bibfield  {journal} {\bibinfo  {journal} {Phys.
  Rev. B}\ }\textbf {\bibinfo {volume} {99}},\ \bibinfo {pages} {104206}
  (\bibinfo {year} {2019}{\natexlab{b}})}\BibitemShut {NoStop}%
\bibitem [{\citenamefont {De~Tomasi}\ \emph {et~al.}(2019)\citenamefont
  {De~Tomasi}, \citenamefont {Hetterich}, \citenamefont {Sala},\ and\
  \citenamefont {Pollmann}}]{GDT_2020_Fock}%
  \BibitemOpen
  \bibfield  {author} {\bibinfo {author} {\bibfnamefont {Giuseppe}\
  \bibnamefont {De~Tomasi}}, \bibinfo {author} {\bibfnamefont {Daniel}\
  \bibnamefont {Hetterich}}, \bibinfo {author} {\bibfnamefont {Pablo}\
  \bibnamefont {Sala}}, \ and\ \bibinfo {author} {\bibfnamefont {Frank}\
  \bibnamefont {Pollmann}},\ }\bibfield  {title} {\enquote {\bibinfo {title}
  {Dynamics of strongly interacting systems: From fock-space fragmentation to
  many-body localization},}\ }\href {\doibase 10.1103/PhysRevB.100.214313}
  {\bibfield  {journal} {\bibinfo  {journal} {Phys. Rev. B}\ }\textbf {\bibinfo
  {volume} {100}},\ \bibinfo {pages} {214313} (\bibinfo {year}
  {2019})}\BibitemShut {NoStop}%
\bibitem [{\citenamefont {Kravtsov}\ \emph {et~al.}(2015)\citenamefont
  {Kravtsov}, \citenamefont {Khaymovich}, \citenamefont {Cuevas},\ and\
  \citenamefont {Amini}}]{Kravtsov_NJP2015}%
  \BibitemOpen
  \bibfield  {author} {\bibinfo {author} {\bibfnamefont {V~E}\ \bibnamefont
  {Kravtsov}}, \bibinfo {author} {\bibfnamefont {I~M}\ \bibnamefont
  {Khaymovich}}, \bibinfo {author} {\bibfnamefont {E}~\bibnamefont {Cuevas}}, \
  and\ \bibinfo {author} {\bibfnamefont {M}~\bibnamefont {Amini}},\ }\bibfield
  {title} {\enquote {\bibinfo {title} {A random matrix model with localization
  and ergodic transitions},}\ }\href
  {http://stacks.iop.org/1367-2630/17/i=12/a=122002} {\bibfield  {journal}
  {\bibinfo  {journal} {New J. Phys.}\ }\textbf {\bibinfo {volume} {17}},\
  \bibinfo {pages} {122002} (\bibinfo {year} {2015})}\BibitemShut {NoStop}%
\bibitem [{\citenamefont {Kravtsov}\ \emph {et~al.}(2020)\citenamefont
  {Kravtsov}, \citenamefont {Khaymovich}, \citenamefont {Altshuler},\ and\
  \citenamefont {Ioffe}}]{LNRP2020_RRG}%
  \BibitemOpen
  \bibfield  {author} {\bibinfo {author} {\bibfnamefont {V.~E.}\ \bibnamefont
  {Kravtsov}}, \bibinfo {author} {\bibfnamefont {I.~M.}\ \bibnamefont
  {Khaymovich}}, \bibinfo {author} {\bibfnamefont {B.~L.}\ \bibnamefont
  {Altshuler}}, \ and\ \bibinfo {author} {\bibfnamefont {L.~B.}\ \bibnamefont
  {Ioffe}},\ }\href@noop {} {\enquote {\bibinfo {title} {Localization
  transition on the random regular graph as an unstable tricritical point in a
  log-normal rosenzweig-porter random matrix ensemble},}\ } (\bibinfo {year}
  {2020}),\ \Eprint {http://arxiv.org/abs/2002.02979} {arXiv:2002.02979}
  \BibitemShut {NoStop}%
\bibitem [{\citenamefont {Khaymovich}\ \emph {et~al.}(2020)\citenamefont
  {Khaymovich}, \citenamefont {Kravtsov}, \citenamefont {Altshuler},\ and\
  \citenamefont {Ioffe}}]{LNRP2020_WE}%
  \BibitemOpen
  \bibfield  {author} {\bibinfo {author} {\bibfnamefont {IM}~\bibnamefont
  {Khaymovich}}, \bibinfo {author} {\bibfnamefont {VE}~\bibnamefont
  {Kravtsov}}, \bibinfo {author} {\bibfnamefont {BL}~\bibnamefont {Altshuler}},
  \ and\ \bibinfo {author} {\bibfnamefont {LB}~\bibnamefont {Ioffe}},\
  }\href@noop {} {\enquote {\bibinfo {title} {Fragile ergodic phases in
  logarithmically-normal rosenzweig-porter model},}\ } (\bibinfo {year}
  {2020}),\ \Eprint {http://arxiv.org/abs/2006.04827} {arXiv:2006.04827}
  \BibitemShut {NoStop}%
\bibitem [{\citenamefont {Beugeling}\ \emph {et~al.}(2015)\citenamefont
  {Beugeling}, \citenamefont {Andreanov},\ and\ \citenamefont
  {Haque}}]{Haque2015PR_vs_Sent}%
  \BibitemOpen
  \bibfield  {author} {\bibinfo {author} {\bibfnamefont {W}~\bibnamefont
  {Beugeling}}, \bibinfo {author} {\bibfnamefont {A}~\bibnamefont {Andreanov}},
  \ and\ \bibinfo {author} {\bibfnamefont {Masudul}\ \bibnamefont {Haque}},\
  }\bibfield  {title} {\enquote {\bibinfo {title} {Global characteristics of
  all eigenstates of local many-body hamiltonians: participation ratio and
  entanglement entropy},}\ }\href {\doibase 10.1088/1742-5468/2015/02/p02002}
  {\bibfield  {journal} {\bibinfo  {journal} {Journal of Statistical Mechanics:
  Theory and Experiment}\ }\textbf {\bibinfo {volume} {2015}},\ \bibinfo
  {pages} {P02002} (\bibinfo {year} {2015})}\BibitemShut {NoStop}%
\bibitem [{\citenamefont {Luitz}\ and\ \citenamefont
  {Bar~Lev}(2016)}]{LuitzBarlev_PRL16}%
  \BibitemOpen
  \bibfield  {author} {\bibinfo {author} {\bibfnamefont {David~J.}\
  \bibnamefont {Luitz}}\ and\ \bibinfo {author} {\bibfnamefont {Yevgeny}\
  \bibnamefont {Bar~Lev}},\ }\bibfield  {title} {\enquote {\bibinfo {title}
  {Anomalous thermalization in ergodic systems},}\ }\href {\doibase
  10.1103/PhysRevLett.117.170404} {\bibfield  {journal} {\bibinfo  {journal}
  {Phys. Rev. Lett.}\ }\textbf {\bibinfo {volume} {117}},\ \bibinfo {pages}
  {170404} (\bibinfo {year} {2016})}\BibitemShut {NoStop}%
\bibitem [{\citenamefont {Beugeling}\ \emph {et~al.}(2018)\citenamefont
  {Beugeling}, \citenamefont {B\"acker}, \citenamefont {Moessner},\ and\
  \citenamefont {Haque}}]{Beugeling_coefficients_PRE2018}%
  \BibitemOpen
  \bibfield  {author} {\bibinfo {author} {\bibfnamefont {Wouter}\ \bibnamefont
  {Beugeling}}, \bibinfo {author} {\bibfnamefont {Arnd}\ \bibnamefont
  {B\"acker}}, \bibinfo {author} {\bibfnamefont {Roderich}\ \bibnamefont
  {Moessner}}, \ and\ \bibinfo {author} {\bibfnamefont {Masudul}\ \bibnamefont
  {Haque}},\ }\bibfield  {title} {\enquote {\bibinfo {title} {Statistical
  properties of eigenstate amplitudes in complex quantum systems},}\ }\href
  {\doibase 10.1103/PhysRevE.98.022204} {\bibfield  {journal} {\bibinfo
  {journal} {Phys. Rev. E}\ }\textbf {\bibinfo {volume} {98}},\ \bibinfo
  {pages} {022204} (\bibinfo {year} {2018})}\BibitemShut {NoStop}%
\bibitem [{\citenamefont {Hamazaki}\ and\ \citenamefont
  {Ueda}(2018)}]{HamazakiUeda_atypicality_PRL2018}%
  \BibitemOpen
  \bibfield  {author} {\bibinfo {author} {\bibfnamefont {Ryusuke}\ \bibnamefont
  {Hamazaki}}\ and\ \bibinfo {author} {\bibfnamefont {Masahito}\ \bibnamefont
  {Ueda}},\ }\bibfield  {title} {\enquote {\bibinfo {title} {Atypicality of
  most few-body observables},}\ }\href {\doibase
  10.1103/PhysRevLett.120.080603} {\bibfield  {journal} {\bibinfo  {journal}
  {Phys. Rev. Lett.}\ }\textbf {\bibinfo {volume} {120}},\ \bibinfo {pages}
  {080603} (\bibinfo {year} {2018})}\BibitemShut {NoStop}%
\bibitem [{\citenamefont {B{\"{a}}cker}\ \emph {et~al.}(2019)\citenamefont
  {B{\"{a}}cker}, \citenamefont {Haque},\ and\ \citenamefont
  {Khaymovich}}]{Baecker2019}%
  \BibitemOpen
  \bibfield  {author} {\bibinfo {author} {\bibfnamefont {Arnd}\ \bibnamefont
  {B{\"{a}}cker}}, \bibinfo {author} {\bibfnamefont {Masudul}\ \bibnamefont
  {Haque}}, \ and\ \bibinfo {author} {\bibfnamefont {Ivan~M}\ \bibnamefont
  {Khaymovich}},\ }\bibfield  {title} {\enquote {\bibinfo {title}
  {{Multifractal dimensions for random matrices, chaotic quantum maps, and
  many-body systems}},}\ }\href {\doibase 10.1103/PhysRevE.100.032117}
  {\bibfield  {journal} {\bibinfo  {journal} {Phys. Rev. E}\ }\textbf {\bibinfo
  {volume} {100}},\ \bibinfo {pages} {032117} (\bibinfo {year}
  {2019})}\BibitemShut {NoStop}%
\bibitem [{\citenamefont {Haque}\ \emph {et~al.}()\citenamefont {Haque},
  \citenamefont {McClarty},\ and\ \citenamefont {Khaymovich}}]{Sent2020_Haque}%
  \BibitemOpen
  \bibfield  {author} {\bibinfo {author} {\bibfnamefont {Masudul}\ \bibnamefont
  {Haque}}, \bibinfo {author} {\bibfnamefont {Paul~A}\ \bibnamefont
  {McClarty}}, \ and\ \bibinfo {author} {\bibfnamefont {Ivan~M}\ \bibnamefont
  {Khaymovich}},\ }\href@noop {} {\enquote {\bibinfo {title} {Entanglement of
  mid-spectrum eigenstates of chaotic many-body systems--deviation from random
  ensembles},}\ }\Eprint {http://arxiv.org/abs/2008.12782} {arXiv:2008.12782}
  \BibitemShut {NoStop}%
\end{thebibliography}%

\end{document}